\documentclass[aps,pra]{revtex4}
\usepackage{graphicx}
\usepackage{colordvi}
\begin{document}

\title{Generation of density waves in dipolar quantum gases by time-periodic modulation of atomic interactions}
\author{B. Kh. Turmanov$^1$, B. B. Baizakov$^1$, F. Kh. Abdullaev$^{1,2}$}
\affiliation{
$^1$ Physical-Technical Institute, Uzbek Academy of Sciences, 100084, Tashkent, Uzbekistan \\
$^2$ Department of Theoretical Physics, National University of
Uzbekistan, Tashkent 100174, Uzbekistan }
\date{\today}

\begin{abstract}

We study the emergence of density waves in dipolar Bose-Einstein
condensates (BEC) when the strength of dipole-dipole atomic
interactions is periodically varied in time. The proposed
theoretical model, based on the evolution of small perturbations of
the background density, allows to compute the growth rate of
instability (gain factor) for arbitrary set of input parameters,
thus to identify the regions of instability against density waves.
We find that among other modes of the system the roton mode is most
effectively excited due to the contribution of sub-harmonics of the
excitation frequency. The frequency of temporal oscillations of
emerging density waves coincides with the half of the driving
frequency, this being the hallmark of the parametric resonance, is
characteristic to Faraday waves. The possibility to create density
waves in dipolar BECs, which can persist after the emergence, has
been demonstrated. The existence of a stationary spatially periodic
solution of the nonlocal Gross-Pitaevskii equation has been
discussed. The effect of three-body atomic interactions, which is
relevant to condensates with increased density, upon the properties
of emerging waves has been analyzed too. Significant modification of
the condensate's excitation spectrum owing to three-body effects is
shown.

\end{abstract}

\maketitle

\section{Introduction}

During the last few years significant new achievements in the field
of dipolar Bose-Einstein condensates (BEC) have been reported, such
as self-bound quantum droplets \cite{ferrier-barbut2016,
schmitt2016}, experimental observation of the quasi-particle called
``roton" \cite{chomaz2018,petter2019} and supersolid phase of matter
\cite{tanzi2019,bottcher2019,chomaz2019,roccuzzo2019,kora2019}.
These exotic phenomena can show up in dipolar quantum gases owing to
a long-range and anisotropic nature of the atomic interactions.
Besides, the dipolar interactions is responsible for a significant
modification of well known phenomena that were observed and studied
in ordinary quantum gases with short-range atomic interactions. For
instance, the spectrum of elementary excitations, features of
modulational instability, interactions between solitons, emergence
and evolution of Faraday waves, and some other phenomena are
substantially different in quantum gases with long-range (dipolar)
atomic interactions as compared to ones with short-range (contact)
interactions. Another notable feature of dipolar BECs is that they
support stable bound states of solitons, so called soliton molecules
\cite{lakomy2012a,baizakov2015,baizakov2019}.

The Faraday instability belongs to extensively studied subjects in
physics. Here we briefly recall that emergence of characteristic
wave patterns on the surface of a vertically vibrating water vessel,
upon reaching certain frequency and amplitude, was observed for the
first time by Michael Faraday \cite{faraday1831}. The Faraday
instability is the classical example of parametric instability,
which occurs when the characteristic coefficients of the nonlinear
system are periodically varied in time. As outcome of such
instability a spatially periodic wave emerges in the medium with
initially uniform properties, which is referred to as a Faraday
wave. To date the Faraday waves are investigated in many areas of
science, including the hydrodynamics, nonlinear optics, chemical
reactions and biological systems, and still represents an active
field of research (see e.g. review articles
\cite{cross1993,nguyen2019}). Faraday waves are extensively studied
also in BECs, both theoretically
\cite{staliunas2002,nikolin2007,nath2010,balaz2014,abdullaev2013,
abdullaev2016,abdullaev2017,abdullaev2019,vudragovic2019} and
experimentally \cite{nguyen2019, engels2007}. An interesting
situation, when the effective atomic interaction in a two-component
BEC periodically changes due to the Rabi oscillations, leading to
Faraday patterns through parametric resonance, has been reported in
a recent paper \cite{chen2019}. This case is special because Faraday
patterns are generated without any modulation of external parameters
of the system. Here the recurrent population imbalance between two
hyperfine states of a binary BEC leads to periodic change of the
strength of effective interaction, which in its turn, triggers the
Faraday instability.

Dipolar BECs represent an excellent system for investigation of
nonlocal effects in quantum gases. The interplay between
nonlinearity and nonlocality clearly shows up in the dynamics of
dipolar BEC by significantly altering its excitation spectrum and
regions of Faraday instability \cite{lakomy2012}. The existence of
roton quasi-particles in dipolar BECs, i.e. rotonization of the
excitation spectrum \cite{chomaz2018}, can be revealed from features
of emerging Faraday patterns. Specifically, the presence of even
shallow roton minimum in the dispersion curve, leads to abrupt
changes in the dependence of pattern size as a function of the
driving frequency \cite{nath2010}, as opposed to monotonic
dependence in roton-free condensates~\cite{engels2007}. Due to high
sensitivity of density patterns to the presence of rotons, dipolar
BECs can represent an excellent system to explore the physics of
rotons.

In certain conditions the effect of higher order nonlinearities may
become essential for consistent interpretation of experimental
results. For instance, important role of three-body atomic
interactions in the formation of quantum droplet patterns in a
$^{164}$Dy condensate, was demonstrated in numerical simulations of
the nonlocal Gross-Pitaevskii equation with cubic-quintic
nonlinearity \cite{xi2016}.

In this paper we study the density waves in dipolar BECs, emerging
and evolving under periodically varying strength of the
dipole-dipole atomic interactions. The contribution of three-body
atomic interactions upon the dynamics of density waves has been
investigated too. Related phenomena in the context of ordinary
BECs with contact interactions were explored in Ref.
\cite{abdullaev2016}. However, as pointed out earlier, the
long-range dipolar atomic interactions drastically modify the
whole dynamics of the condensate, including the features of
density waves. Therefore, we can expect to observe new facets of
density waves in dipolar BECs when three-body atomic interactions
are taken into account. The remaining part of the paper is
organized as follows. In the next Sec. II we introduce the
mathematical model and derive a Mathieu type equation to address
the wave instability in dipolar BEC. In Sec. III we discuss the
results of numerical simulations. Main results of this work is
summarized in the last Sec. IV.

\section{The model and analysis of wave instability}

Below we consider a dipolar BEC with two- and three- body atomic
interactions. We assume that atomic dipoles are oriented along the
$x$- axis of the trap. The dynamics of the condensate is described
by the following 3D Gross-Pitaevskii equation (GPE):
\begin{equation} \label{gpe3d}
i \hbar\frac{\partial \Psi(\vec{r},t)}{\partial
t}=\left[-\frac{\hbar^2}{2 m} \vec{\nabla}^2 +
V(\vec{r})+g_{1}|\Psi(\vec{r},t)|^2 + g_{2} |\Psi(\vec{r},t)|^4
+\frac{C_{d}}{4 \pi} \int_{-\infty}^{\infty}\frac{1-3
\frac{(x-x')^2}{|\vec{r}-\vec{r'}|^2}}{|\vec{r}-\vec{r'}|^3} \,
|\Psi(\vec{r'},t)|^2 d\vec{r'} \right] \, \Psi(\vec{r},t),
\end{equation}
where $\Psi(\vec{r},t)$ is the mean-field wave function of the
condensate, normalized to the number of atoms $N =
\int_{-\infty}^{+\infty} |\Psi(\vec{r})|^2 d\vec{r}$, $m$ is the
atomic mass. The BEC is trapped in external harmonic potential
$V(\vec{r})= m (\omega_{x}^2x^2 + \omega_{y}^2y^2 +
\omega_{z}^2z^2)/2$. The strengths of two-body and three-body atomic
interactions are represented by coefficients $g_{1}=4 \pi \hbar^2
a_{s}/m$ and $g_{2}$, respectively. The dipolar interactions are
expressed through the parameter $C_{d}=\mu_{0}\mu^2$ or
$C_{d}=d^2/\varepsilon_{0}$, depending on whether the atoms possess
a magnetic ($\mu$) or electric ($d$) dipole moment, where $\mu_0$,
$\varepsilon_{0}$ are the permeability and the permittivity of
vacuum, respectively. When the effective length of magnetic dipolar
interactions $a_{dd}=\mu_0 \mu^2 m/12 \pi \hbar^2$ is greater than
the length of $s$-wave contact interactions $a_s$, the condensate is
said to be in the dipolar dominated regime ($a_{dd} > a_s$).

To simplify the analysis, we consider a 1D geometry which assumes
a strong confinement in the radial direction $\omega_{\bot} \gg
\omega_{x}$. In this case the transverse (in $y-z$ plane) dynamics
of the condensate is frozen to the ground state $\phi(y,z) =
\left( 1/\sqrt{\pi}a_{\bot} \right) \, \exp\left(-(y^2 + z^2)/2
a_{\bot}^2 \right)$ of the 2D harmonic potential with $a_{\bot} =
\sqrt{\hbar/m \omega_{\bot}}$ being the harmonic oscillator
length. Consequently, the wave function can be factorized as
$\Psi(\vec{r},t)=\phi(y,z)\,\psi(x,t)$. Then integration over
transverse coordinates yields the following 1D GPE
\begin{eqnarray}
  && i\hbar \frac{\partial \psi}{\partial t}
   = -\frac{\hbar^2}{2m}\frac{\partial^2 \psi}{\partial x^2} +
    \frac{m\omega_{x}^2x^2}{2}\, \psi + \frac{g_{1}}{2 \pi
    a_{\bot}^2} |\psi|^2\psi+
\nonumber \\
   && \frac{g_{2}}{3\pi^2 a_{\bot}^4} |\psi|^4
    \psi - \frac{C_{d}}{2 \pi a_{\bot}^3} \psi
    \int_{-\infty}^{\infty}
    R\left(\frac{x-x'}{a_{\bot}}\right) |\psi(x',t)|^2\,dx',
\label{gpe2}
\end{eqnarray}
where $R(x) = \sqrt{\pi} \exp{\left(x^2\right)} \left( 1+2
x^2\right) \textrm{erfc} \left(|x| \right)- 2|x| $ is the response
function characterizing the nonlocal feature of the dipolar BEC
\cite{sinha2007}.

Next, by introducing new variables $t \rightarrow \omega_{\bot}
t$, $x \rightarrow x/a_{\bot}$ and $\psi \rightarrow \sqrt{2
|a_s|}\psi$, we reduce the Eq. (\ref{gpe2}) into following
dimensionless form
\begin{equation}
\label{gpe}
  i\psi_t+\frac{1}{2}\psi_{xx}+q|\psi|^2\psi + p|\psi|^4\psi +
  g\; \psi(x,t)\int_{-\infty }^{+\infty} R(|x-x'|)\ |\psi
  (x',t)|^2 \,dx' =0,
\end{equation}
where for the sake of clarity we denoted the dimensionless
coefficient of attractive cubic term as $q = a_s/a_{bg}$, with
$a_s$, $a_{bg}$ being the $s$-wave scattering length and its
background value. The coefficients of quintic and dipolar terms have
the form $p = g_2 m^2\omega_{\bot}/12 \pi^2 \hbar^3 a_s^2$,
$g=C_{d}/(4\pi\hbar^2 a_s/m)$. In addition, we have excluded the
weak axial trap potential ($\omega_x \ll \omega_{\bot}$) and adopted
periodic boundary conditions for the wave function. This equation is
valid as long as the chemical potential of the condensate is much
less than the energy of radial excitations $\mu/\hbar \omega_{\bot}
= 2 n_{1D} a_s |1-\varepsilon_{d}| \ll 1$, where
$\varepsilon_{d}=a_{dd}/a_s$ is the ratio between strengths of
dipolar and contact interactions, $n_{1D}$ is the one-dimensional
density of the condensate. Both the dipolar and contact interaction
coefficients are tunable by external magnetic fields
\cite{giovanazzi2002,tang2018,chin2010}.

The density waves in BEC can be induced by periodic modulation of
the parameters $q$, $p$ or $g$ in Eq. (\ref{gpe}). Below we assume
variable dipolar interactions with amplitude $\alpha$ and frequency
$\omega$
\begin{equation}\label{gt}
g(t)=g_0 [1 + 2 \alpha \cos(2\omega t)].
\end{equation}
In the absence of modulations ($\alpha=0$), the condensate remains
in a uniform state with constant density. This state is described
by the plane wave solution of Eq.~(\ref{gpe}) with amplitude $A$
and phase $\theta$
\begin{equation}\label{pw}
\psi = A e^{i \theta(t)}, \quad \theta(t) = A^2(q+g_0 K + p A^2)t,
\quad K=\int_{-\infty}^{\infty} R(|x|)\,dx.
\end{equation}
To study the emergence and dynamics of density waves, we introduce
weak perturbation of the uniform state given by Eq. (\ref{pw}), and
look for a solution in the form
\begin{equation}
\psi = (A + \delta\psi)e^{i\theta}, \quad \delta\psi \ll A.
\label{modpw}
\end{equation}
The equation for the perturbation is obtained by substituting Eq.
(\ref{modpw}) into Eq. (\ref{gpe}), and keeping only the linear
terms on $\delta \psi(x,t)$
\begin{eqnarray}
i\delta\psi_t + \frac{1}{2}\delta\psi_{xx} + A^2(q + 2pA^2)
(\delta\psi + \delta\psi^{\ast}) +
\nonumber\\
g A^2\int_{-\infty}^{\infty}
R(|x-x'|)[\delta\psi(x',t)+\delta\psi^{\ast}(x',t)]\,dx' = 0.
\end{eqnarray}
By representing $\delta \psi = u + iv$ and performing a Fourier
transform, we obtain for the Fourier component \\
$\bar{u}(k,t)=\int_{-\infty}^{\infty} u(x,t) \,e^{i k x}\, dx$ the
Mathieu type equation
\begin{equation}\label{mathieu}
\bar{u}_{tt} + [\Omega^{2}(k) - b  \cos(2\omega t)] \bar{u}=0,
\end{equation}
where
\begin{equation}\label{Omega2}
\Omega^2(k) = \frac{k^2}{4}[k^2 - 4 A^2 (q+2 A^2 p + g_0
\bar{R}(k)],
\end{equation}
\begin{equation}\label{b}
b = 2 A^2 \alpha g_0 k^2 \bar{R}(k),
\end{equation}
\begin{equation}
\bar{R}(k)= 2 \left[1 + \frac{k^2}{4} e^{k^2/4}
{\rm Ei}(-\frac{k^2}{4}) \right], \qquad \mbox{with} \qquad {\rm
Ei}(z)=-\int_{-z}^{\infty} \frac{e^{-t}}{t}\,dt
\end{equation}
being the exponential integral function \cite{abramowitz}.

Equation~(\ref{mathieu}) for the Fourier component of the
perturbation is equivalent to that of the parametrically excited
oscillator with an internal frequency $\Omega(k)$. Similar equation
in the context of Faraday waves in dipolar BEC was derived in
\cite{lakomy2012}. In our setting Eq. (\ref{mathieu}) extends the
results of that work by taking into regard the effect of three-body
atomic interactions.

\begin{figure}[htb]
\centerline{{\large \qquad a)} \hspace{4cm} {\large b)} }
\centerline{
\includegraphics[width=4cm,height=4cm,clip]{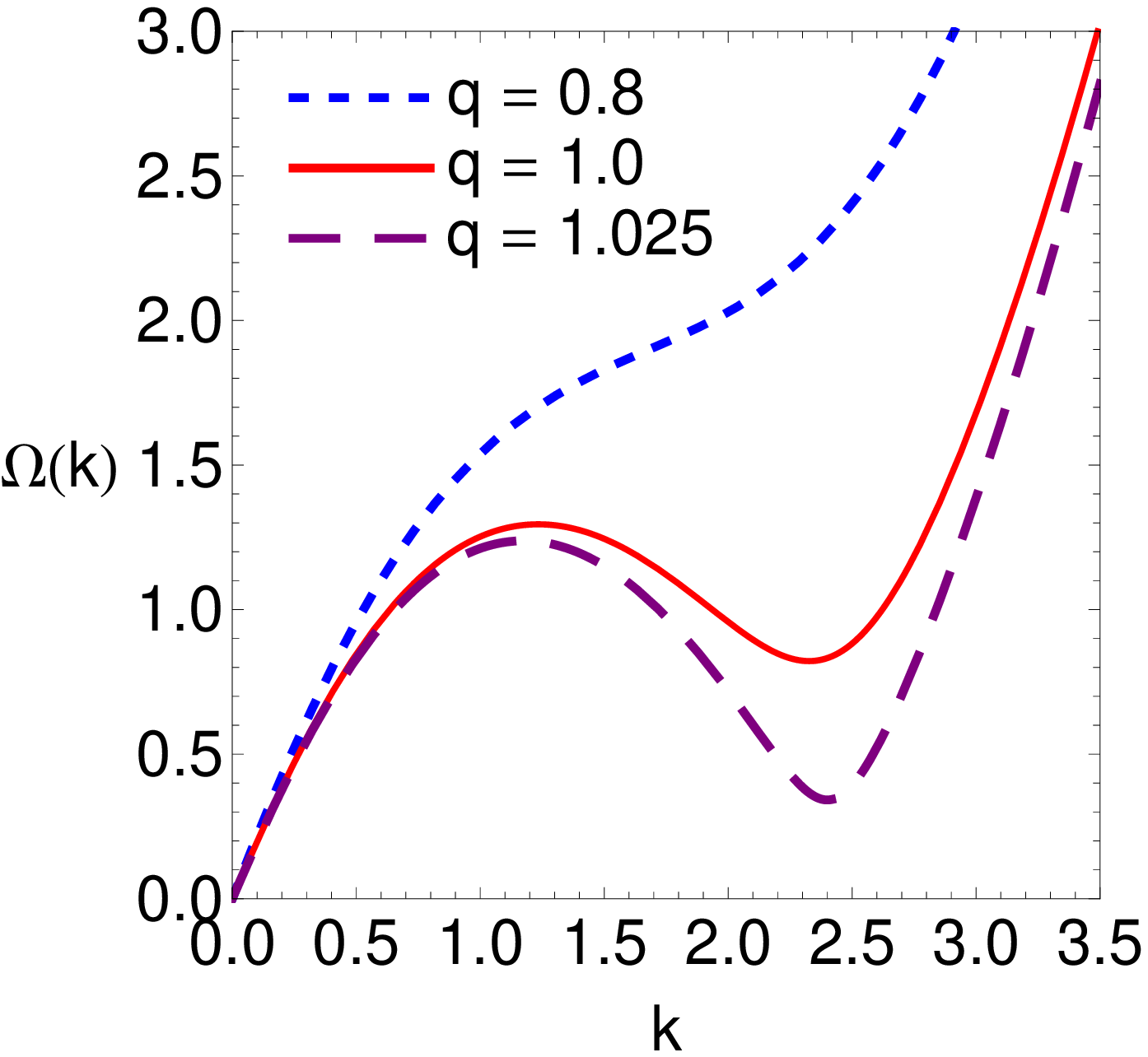}\quad
\includegraphics[width=4cm,height=4cm,clip]{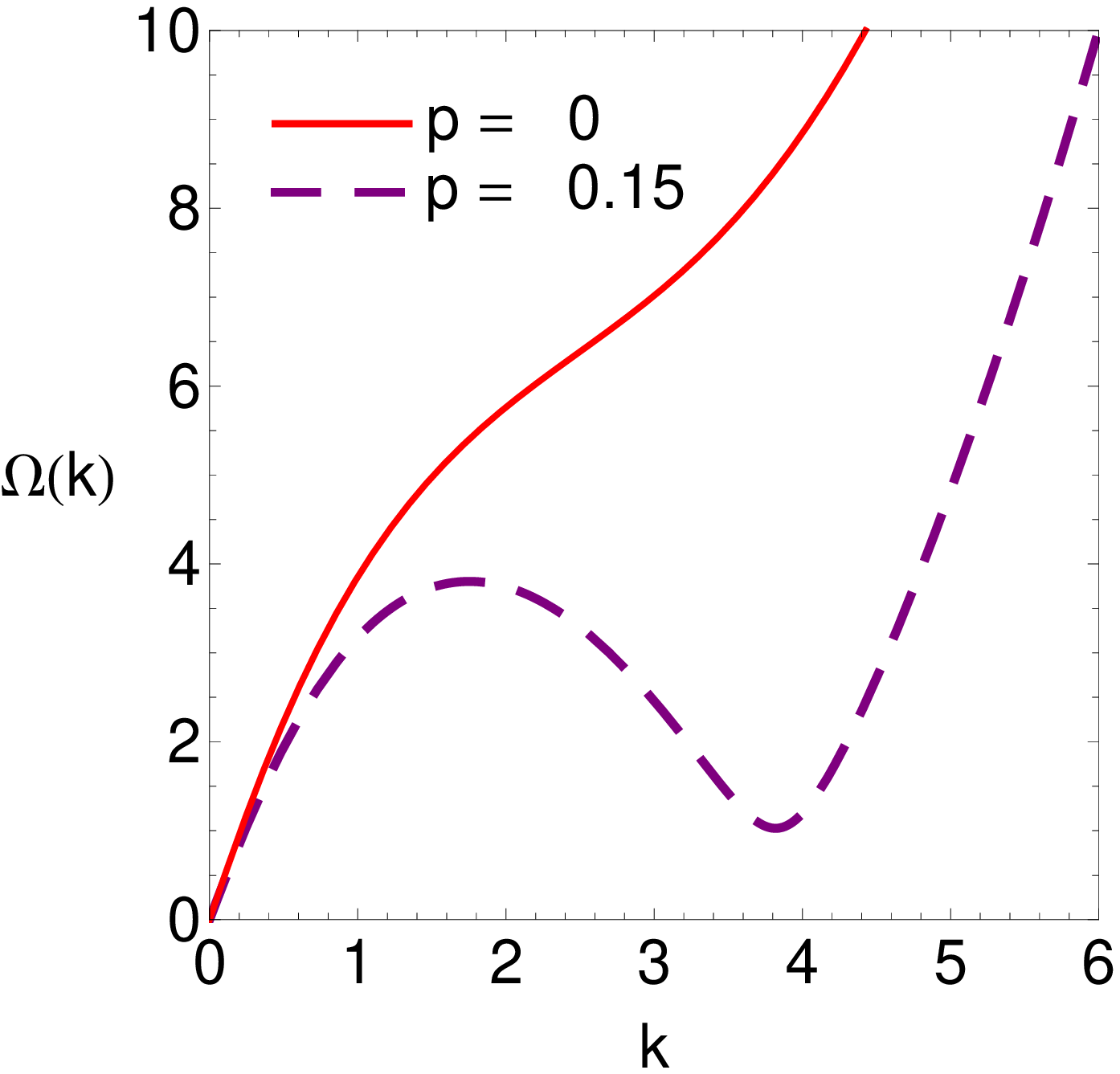}}
\caption{(Color online) The excitation spectrum of a dipolar BEC
according to Eq. (\ref{Omega2}) for different sets of input
parameters. (a) The dispersion curve in absence of the quintic term
($p=0$) features a local minimum at finite momentum $k_{rot} \simeq
2.35$, which indicates to presence of roton quasi-particles in the
condensate (red solid line). If the strength of contact interactions
is slightly reduced ($q=0.8$) or increased ($q=1.025$), the roton
minimum disappears (blue dashed line) or deepens (purple dashed
line), respectively. (b) A small attractive quintic nonlinearity may
induce a roton minimum (purple dashed line) in a roton-free
condensate (red solid line). Parameter values: $A=2$, $q=1$,
$g_0=-1$ (a), $g_0=-3.5$ (b). } \label{fig1}
\end{figure}

The spectrum of elementary excitations defines the basic properties
of the condensed matter. For dipolar BEC it is represented by
dispersion relation Eq. (\ref{Omega2}), which is illustrated in
Fig.~\ref{fig1}.
A remarkable property of dipolar BEC is that, it supports the roton
quasi-particles. A peculiar shape of the dispersion curve with a
local minimum at finite momentum $k_{rot}$, is the hallmark of the
roton mode (red curve in Fig.~\ref{fig1}a). Originally, the
existence of rotons in quantum fluids was postulated by L. Landau in
1941 to explain the properties of superfluid $^4$He
\cite{landau1941}, and later they were observed in neutron
scattering experiments \cite{henshaw1961}. The presence of rotons in
dipolar quantum gases was theoretically predicted in Ref.
\cite{santos2003} and they had been observed in recent experiments
\cite{chomaz2018,petter2019}. Below we shall refer to condensates,
whose excitation spectra possess (do not possess) a local minimum as
rotonized (roton-free) condensates.

Rotons are responsible for the tendency of a system to establish a
local order, with typical length scale $\sim 1/k_{rot}$, which is
equal  approximately to the inter-particle distance. While in
strongly interacting systems, such as superfluid $^4$He, this indeed
holds true, in dilute quantum gases the atomic correlations are due
to the long-range dipolar interactions with a length scale, greatly
exceeding the inter-particle distances. As a consequence of the
roton mode, a long-lived super-solid phase in dipolar BEC can emerge
\cite{tanzi2019}. The quintic nonlinearity, designated by a
coefficient $p$ in Eq. (\ref{gpe}), can significantly modify the
dispersion curve of a dipolar BEC. Even relatively small quintic
nonlinearity may lead to emergence of a roton minimum in a roton
free condensate, as shown in Fig. \ref{fig1}b.

Time-periodic modulation of the coefficient of nonlinearity may lead
to generation of different kinds of elementary excitations
(quasi-particles) in the system, such as phonons, rotons, resonance
waves, Faraday waves, etc. Which kinds of these waves are
predominantly generated depends primarily on the properties of the
condensate, and external parameters, such as the amplitude and
frequency of the perturbation. In some regions of the parameter
space, several types of waves may be simultaneously generated. All
above mentioned types of excitations reveal themselves as density
modulations in the condensate, thus below we refer to them as
density waves.

\section{Numerical simulations}

For numerical simulations of the nonlocal GPE (\ref{gpe}) we use the
split-step fast Fourier transform method \cite{agrawal} with 2048
Fourier modes and periodic boundary conditions. The integration
domain was selected sufficiently large $x~\in~[-40 \pi, 40 \pi]$, to
accommodate $\sim 100 $ periods of emerging waves on average, and
the time step was $\Delta t = 0.0005$. To evaluate the integral term
of the equation the convolution theorem has been employed. To
minimize disturbances, caused by a possible mismatch between the
domain length and integer number of periods of emerging waves, we
consider only the small amplitude limit. As initial perturbation of
the background amplitude we use random numbers ${\cal R}(x)$
uniformly distributed in the interval [-1,1]: $\psi(x_i,0)= A +
\sigma {\cal R}(x_i)$, with $\sigma=0.001$ being the strength of the
perturbations.

When the amplitude of a density wave starts to increase rapidly due
to a parametric resonance, typically exceeding one percent of the
background amplitude, we interrupt the numerical simulation and
analyze the resulting spatial pattern. As a rule, the spatial
pattern represents a superposition of waves with different spatial
periods and shows up as a high frequency component modulated by a
low frequency envelope. Modulation of the amplitude originates from
constructive and destructive interference of emerging density waves
with different periods.

An essential part of numerical simulations is concerned with
revealing the instability domains of Eq. (\ref{mathieu}), which can
be represented as a system of two first order equations
\begin{equation}\label{uv}
{\bar u}_t=-{\bar v}, \quad {\bar v}_t=f(t)\, {\bar u}, \quad
f(t)=\Omega^{2}(k) - b  \cos(2\omega t).
\end{equation}
Since the coefficient $f(t)$ is a periodic function with a minimal
period $T=\pi/\omega$, one can use the Floquet theory to analyze the
stability properties of this system \cite{jordan2007}. To this end,
a fundamental solution matrix out of two solution vectors of Eq.
(\ref{uv}) \{${\bar u}_1(t),{\bar v}_1(t)$\}, \{${\bar u}_2(t),{\bar
v}_2(t)$\} is constructed, which satisfies the initial conditions
\begin{equation}\label{ic}
\left[
  \begin{array}{c}
    u_1(0) \\
    v_1(0) \\
  \end{array}
\right] = \left[
  \begin{array}{c}
    1 \\
    0 \\
  \end{array}
\right], \qquad
\left[
  \begin{array}{c}
    u_2(0) \\
    v_2(0) \\
  \end{array}
\right] = \left[
  \begin{array}{c}
    0 \\
    1 \\
  \end{array}
\right].
\end{equation}
Next, we produce the matrix $C$, which is the fundamental solution
matrix, evaluated at time $T$
\begin{equation}\label{c}
C=
\left[
  \begin{array}{cc}
    u_1(T) & u_2(T) \\
    v_1(T) & v_2(T) \\
  \end{array}
\right].
\end{equation}
According to Floquet's theory \cite{jordan2007}, stability of
solutions to Eq. (\ref{uv}) is determined by the eigenvalues of
the matrix $C$
\begin{equation}\label{evals}
\lambda_{1,2} = \frac{t_c \pm \sqrt{t_c^2-4\, d_c}}{2},
\end{equation}
where $t_c=u_1(T)+v_2(T)$, $d_c=u_1(T)\,v_2(T) - u_2(T)\, v_1(T)$
are the trace and determinant of the matrix $C$, respectively. It
is easy to show using Eqs. (\ref{uv}) and (\ref{ic}) that the
determinant of the fundamental solution matrix is equal to unity
at any time $d_c(t) = d_c(0) = d_c(T) = 1$, therefore
$\lambda_{1,2}=(t_c \pm \sqrt{t_c^2 - 4})/2$. Since the product of
the two eigenvalues is $\lambda_1 \cdot \lambda_2 = 1$, the
instability sets in if either of the eigenvalues has modulus
larger than unity (because the other one is less than unity).
Therefore, the instability develops in all cases when $|t_c| >2$.
If $|t_c|<2$, then Eq. (\ref{evals}) has two complex conjugate
roots, both lying on the unit circle, since their product is
unity. This situation corresponds to stable quasi-periodic motion.
In particular case $t_c = 2$ ($t_c = -2$) the motion is stable
with period $T$ ($2 T$).

An important quantity, relevant to the unstable solution of
Eq.(\ref{uv}) is called a characteristic exponent of the system,
or gain factor (G), and is defined as follows
\begin{equation}\label{gain}
e^{G\,T} = \lambda, \qquad G = \frac{1}{T} Ln(\lambda),
\end{equation}
where the principal value of the logarithm is assumed. The amplitude
of the unstable wave, once emerged, grows exponentially with time
$|\delta \psi| \sim e^{G\,t}$.

\subsection{The domains of instability and features of density waves}

The proposed theoretical approach, expressed through
Eqs.~(\ref{uv})-(\ref{evals}), allows to identify the domains of
instability in the parameter space against generation of density
waves in dipolar BECs. The growth rate of instability, also known as
gain factor, given by Eq. (\ref{gain}), can be considered as a
function of system's parameters $G=F(A,q,g_0,p,\alpha,k,\omega)$,
and its numerical value can be calculated for arbitrary set of these
parameters using the Floquet's theory.

\begin{figure}[htb]
\centerline{\hspace{0.5cm} a) \hspace{4cm} b) \hspace{4cm}  c) }
\centerline{
\includegraphics[width=4cm,height=3cm,clip]{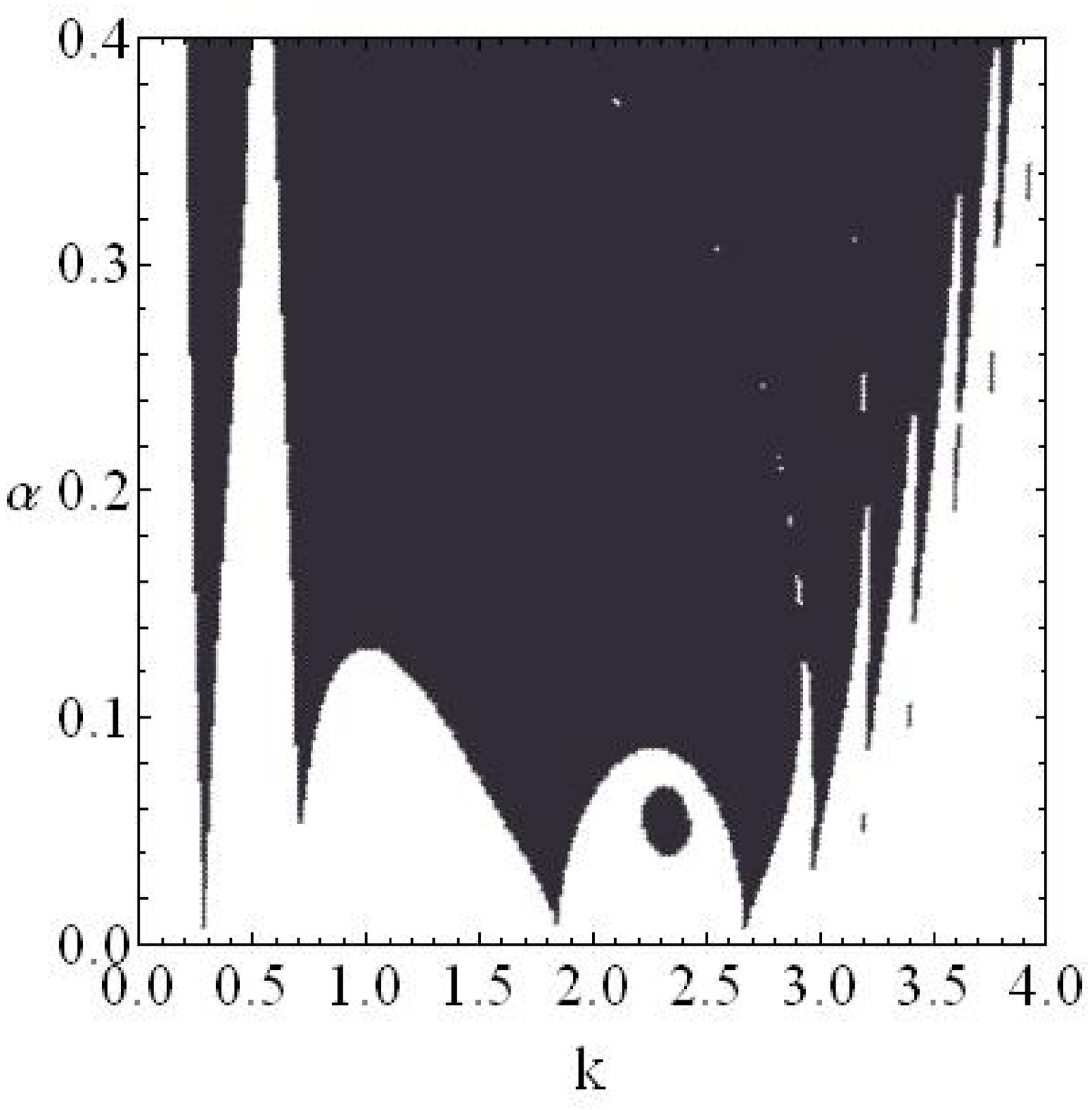}\qquad
\includegraphics[width=4cm,height=3cm,clip]{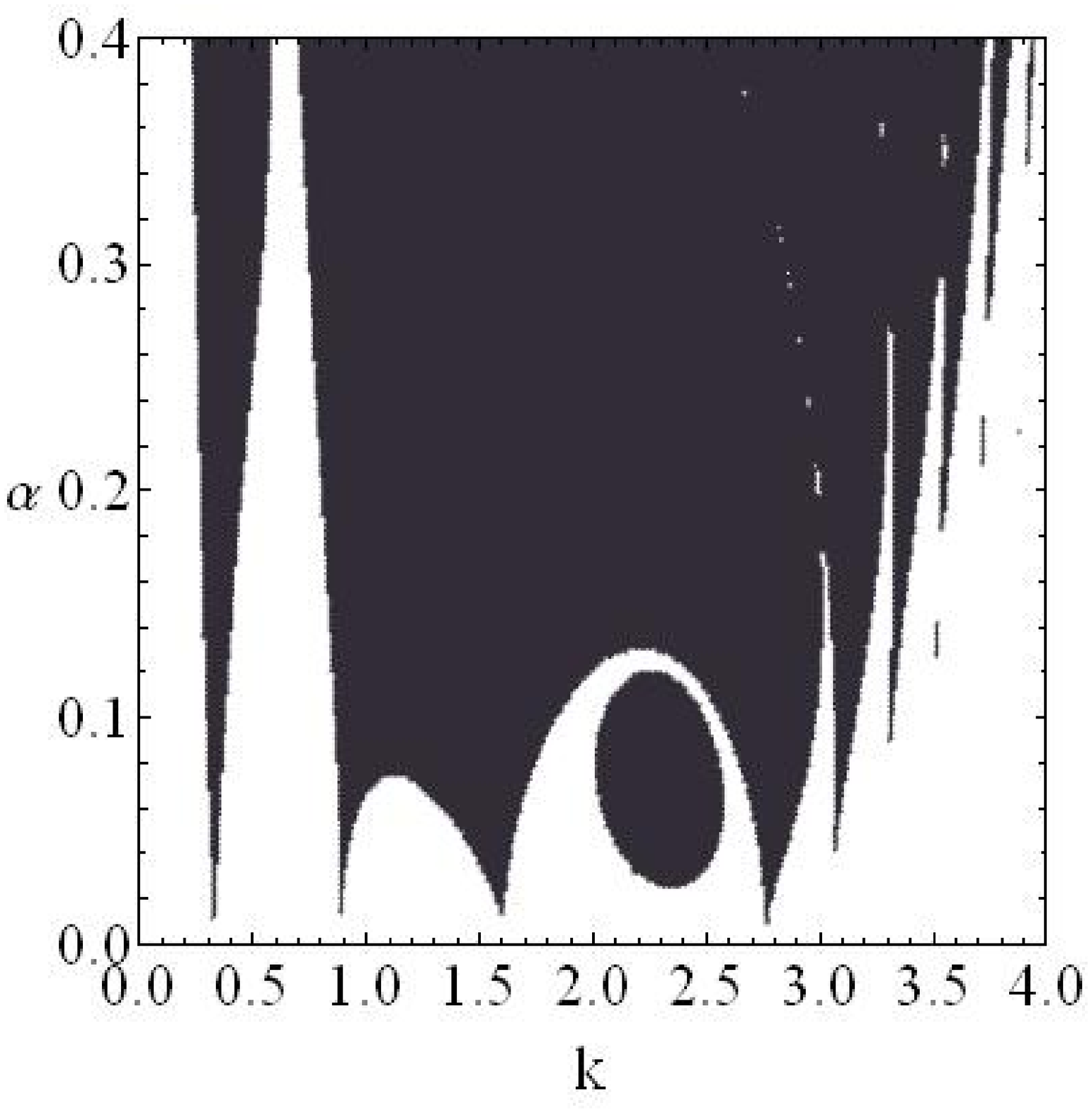}\qquad
\includegraphics[width=4cm,height=3cm,clip]{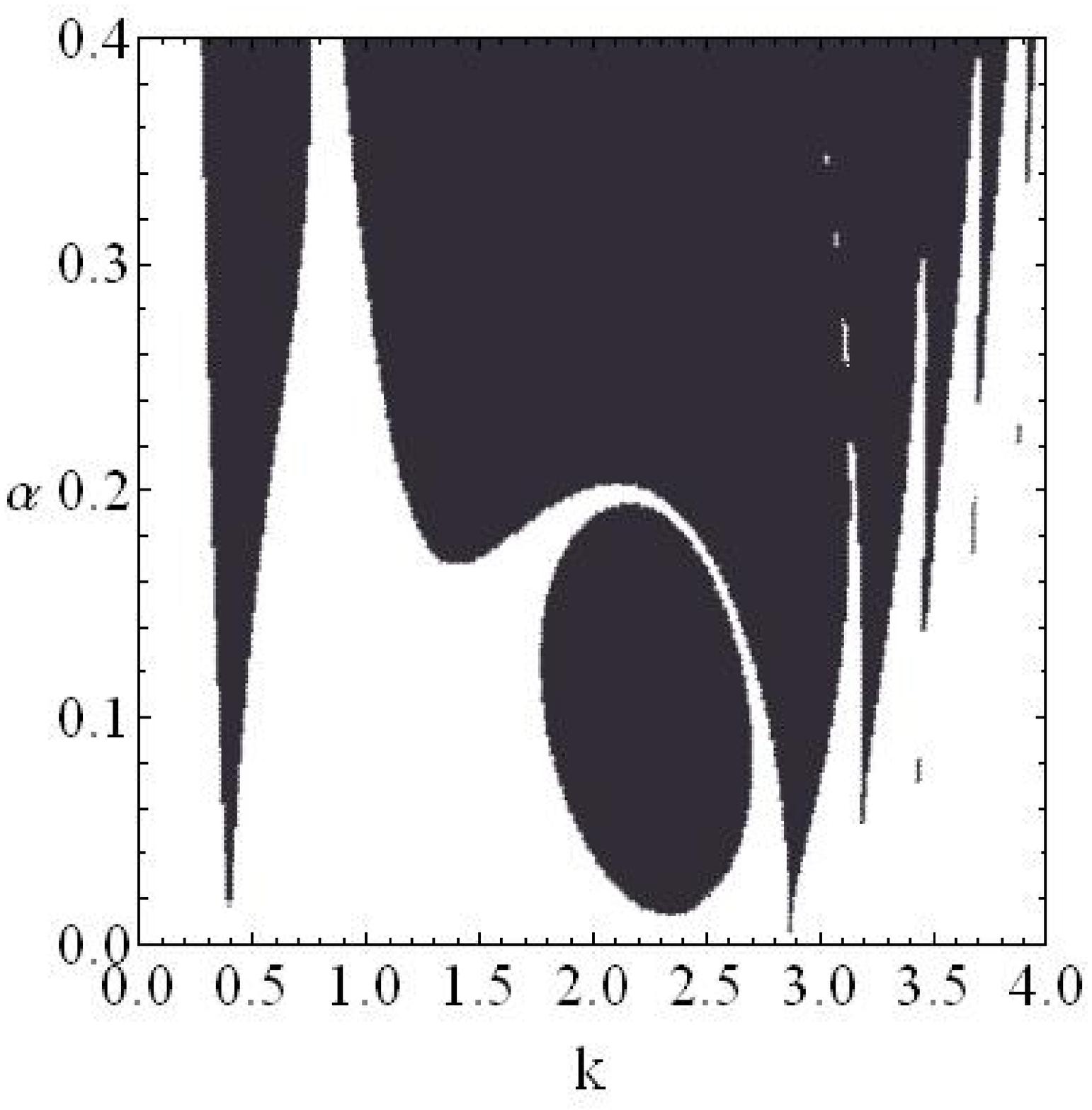}}
\centerline{
\includegraphics[width=4cm,height=3cm,clip]{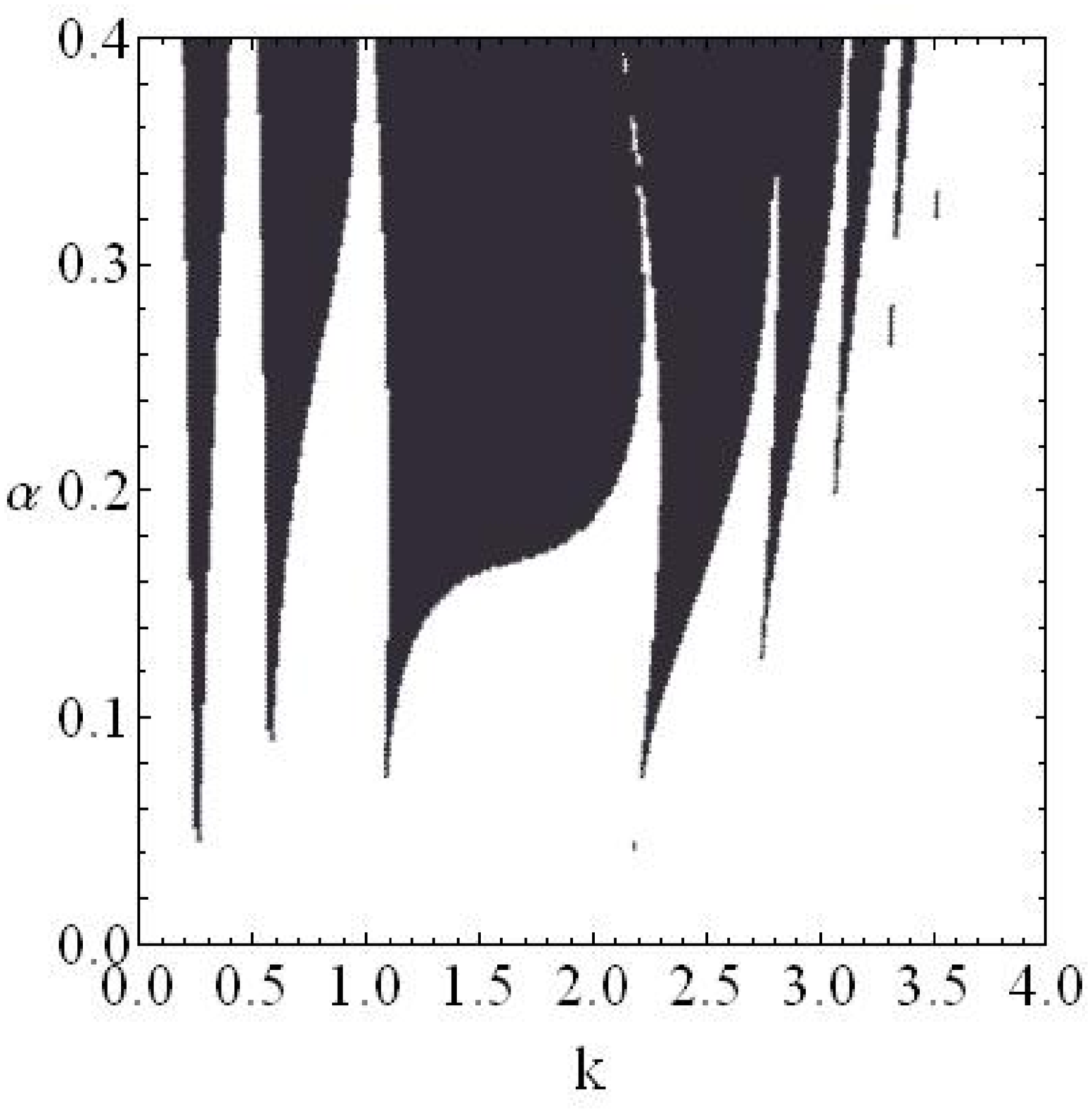}\qquad
\includegraphics[width=4cm,height=3cm,clip]{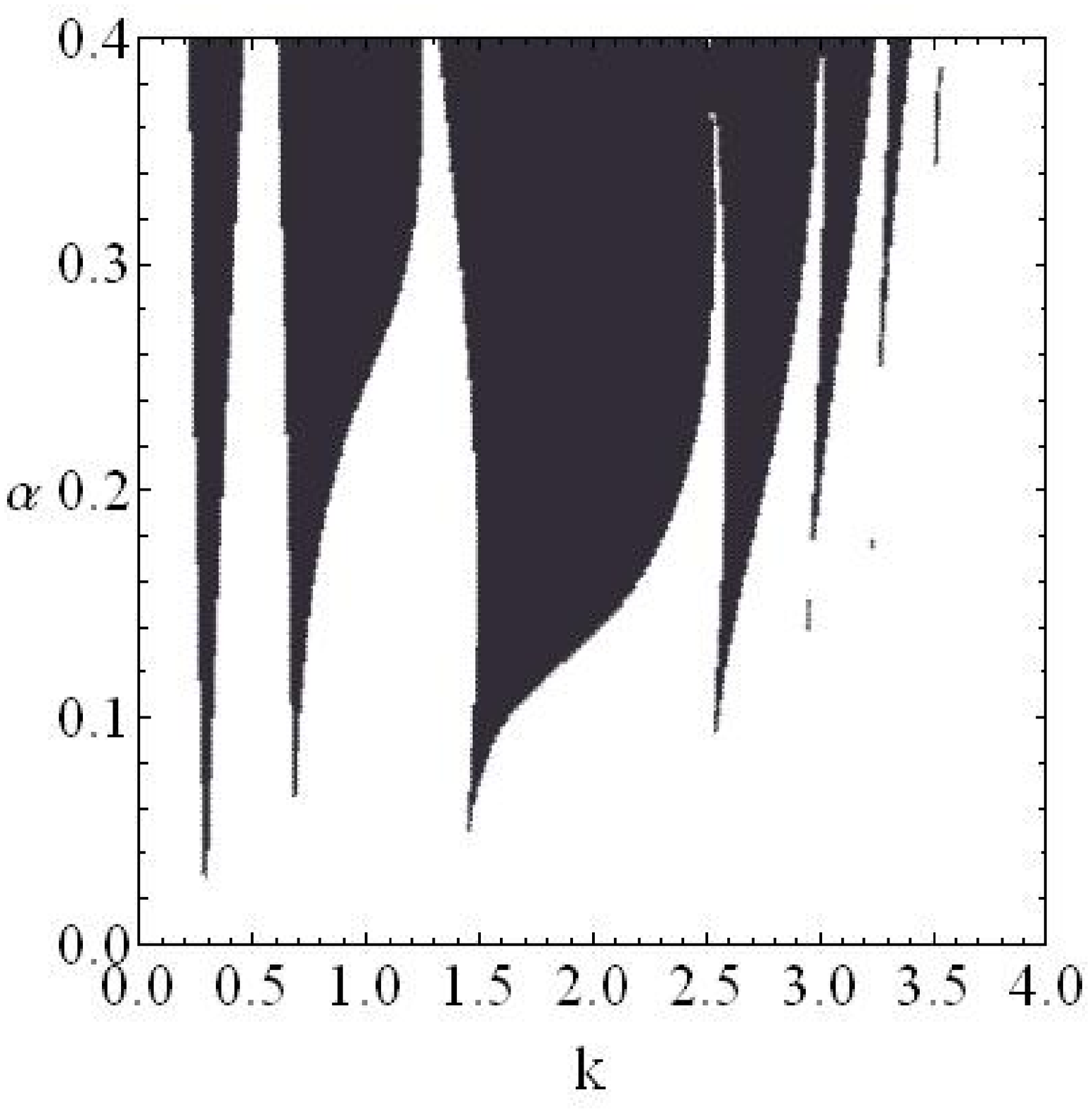}\qquad
\includegraphics[width=4cm,height=3cm,clip]{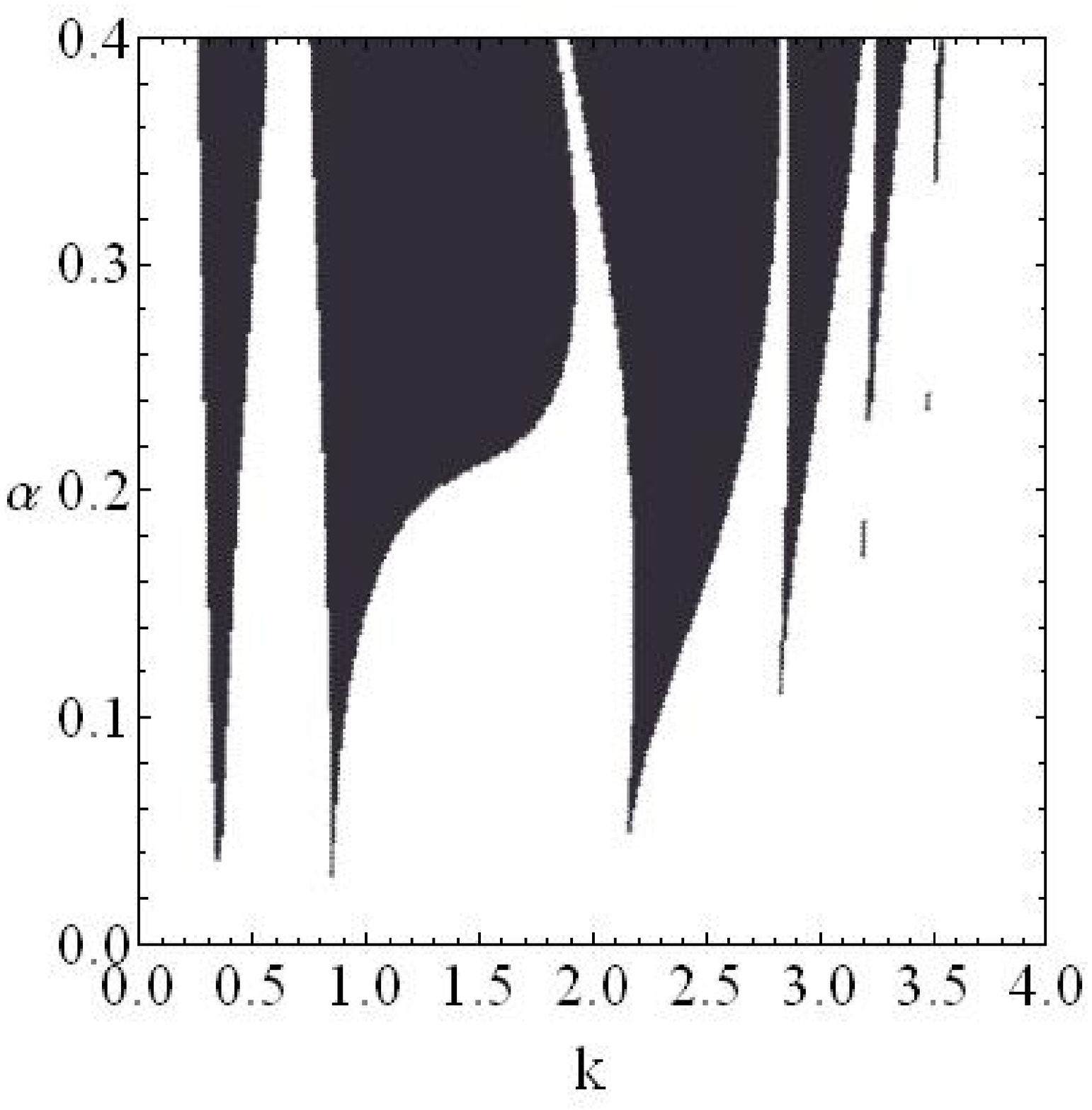}}
\caption{(Color online) Upper panels: The domains of instability
(dark regions) of Eq. (\ref{mathieu}) when the roton mode in the
excitation spectrum of the condensate is present. Pronounced island
of instability appears around $k_{rot}=2.35$ and expands as the
driving frequency increases: $\omega = 0.53$~(a); \ $\omega = 0.6$
(b); \ $\omega = 0.7$ (c). Lower panels: The island of instability
disappears when the the roton mode is eliminated from the spectrum
of excitations (see Fig. \ref{fig1}) by reducing the strength of
contact interactions from $q=1$ to $q=0.8$.} \label{fig2}
\end{figure}

In Fig. \ref{fig2} we show the domains of instability (dark regions)
in the space of two parameters: the wave vector of emerging density
waves ($k$), and the strength of modulation of the dipolar
interactions ($\alpha$), at fixed values of other parameters.
The upper row in this figure corresponds to the case, when the roton
mode in the dispersion curve is present (red line in Fig.
\ref{fig1}a). Conspicuous island of instability appears around
$k_{rot} = 2.35$ at parameter values $\alpha = 0.055, \ \omega =
0.52$, and expands in size as the modulation frequency increases.
This island of instability disappears if the roton mode is
eliminated by reducing the strength of contact interactions from
$q=1$ to $q=0.8$ (blue dashed line in Fig.~\ref{fig1}a), which is
shown in the lower panels of Fig. \ref{fig2}. The island of
instability does not appear also in other settings without the roton
mode, for instance, due to contribution of a quintic nonlinearity in
Eq.~(\ref{Omega2}). These observations suggest, that the revealed
island of instability corresponds to excitation of the roton mode in
dipolar BECs, subject to periodic variation of the strength of
dipolar interactions. It should be noted, that similar island of
instability was found in numerical simulations of the 2D dipolar BEC
\cite{nath2010}, however its relevance to the roton mode was not
explicitly commented there. A qualitative difference between the
instability domains of rotonized and roton-free dipolar condensates
is clearly evident from the upper and lower panels of Fig.
\ref{fig2}.

\begin{figure}[htb]
\centerline{{\large \qquad a)} \hspace{4cm} {\large b)} }
\centerline{
\includegraphics[width=4cm,height=4cm,clip]{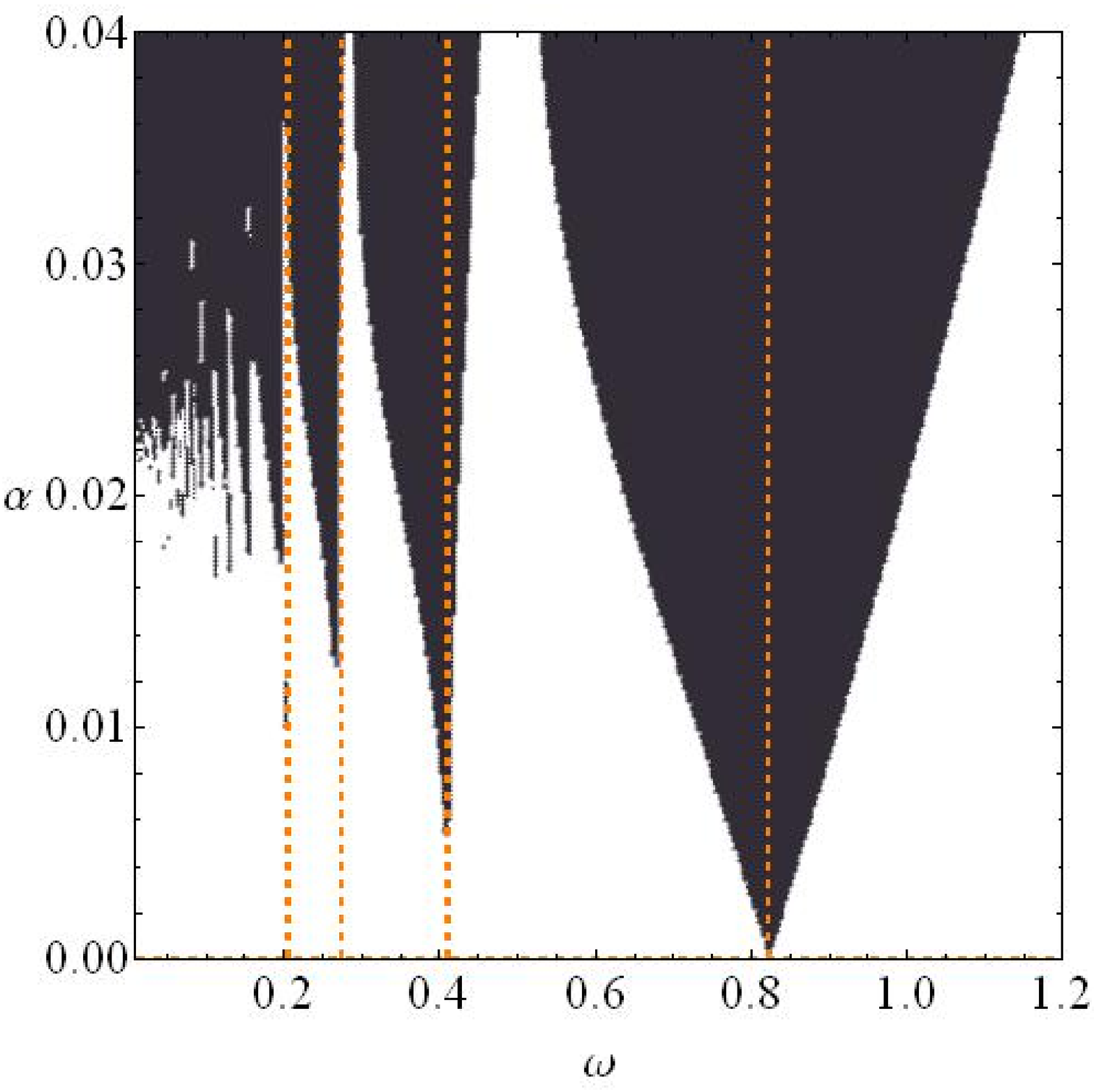}\quad
\includegraphics[width=4cm,height=4cm,clip]{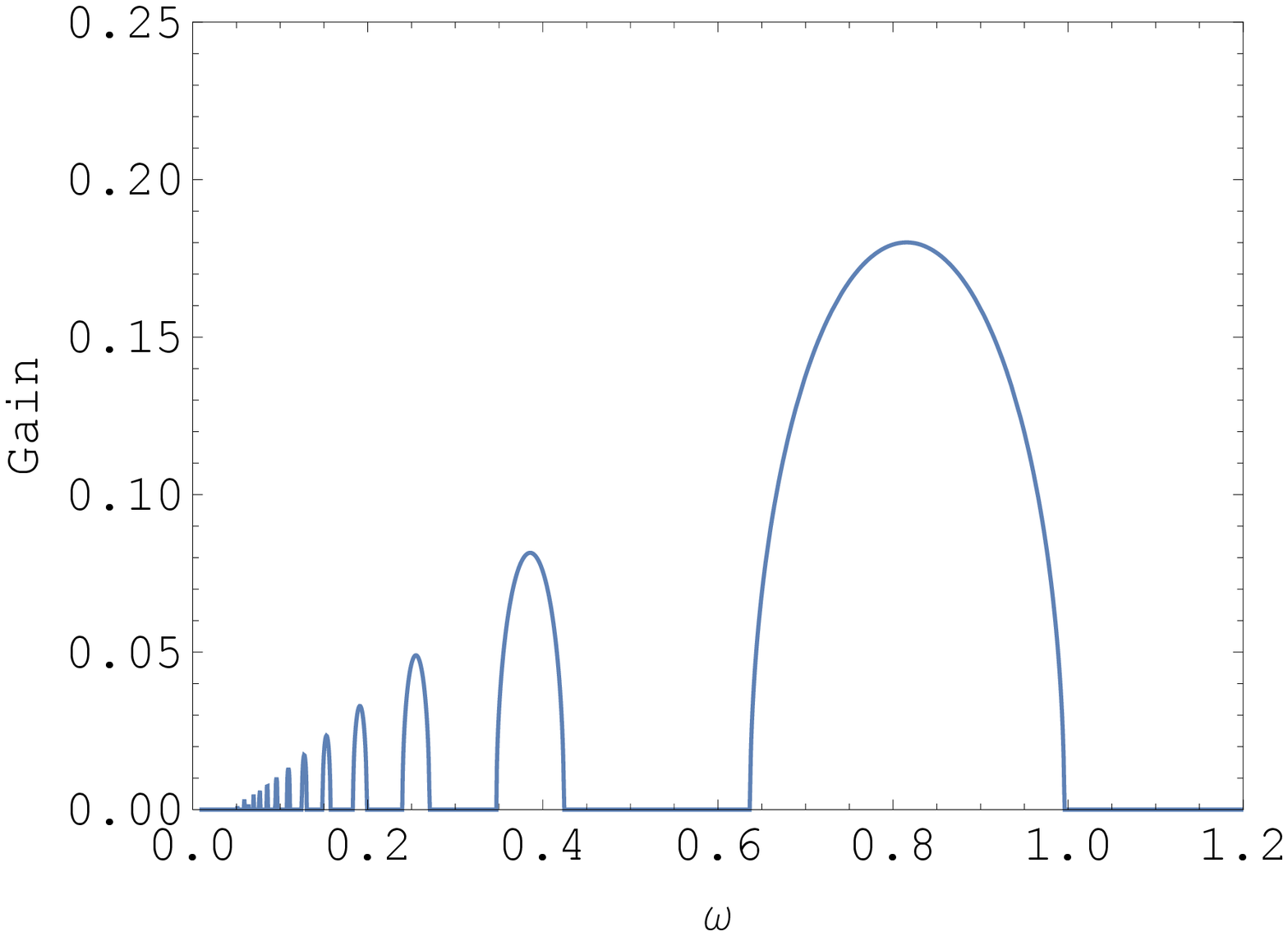}}
\centerline{{\large \qquad c)} \hspace{4cm} {\large d)} }
\centerline{
\includegraphics[width=4cm,height=4cm,clip]{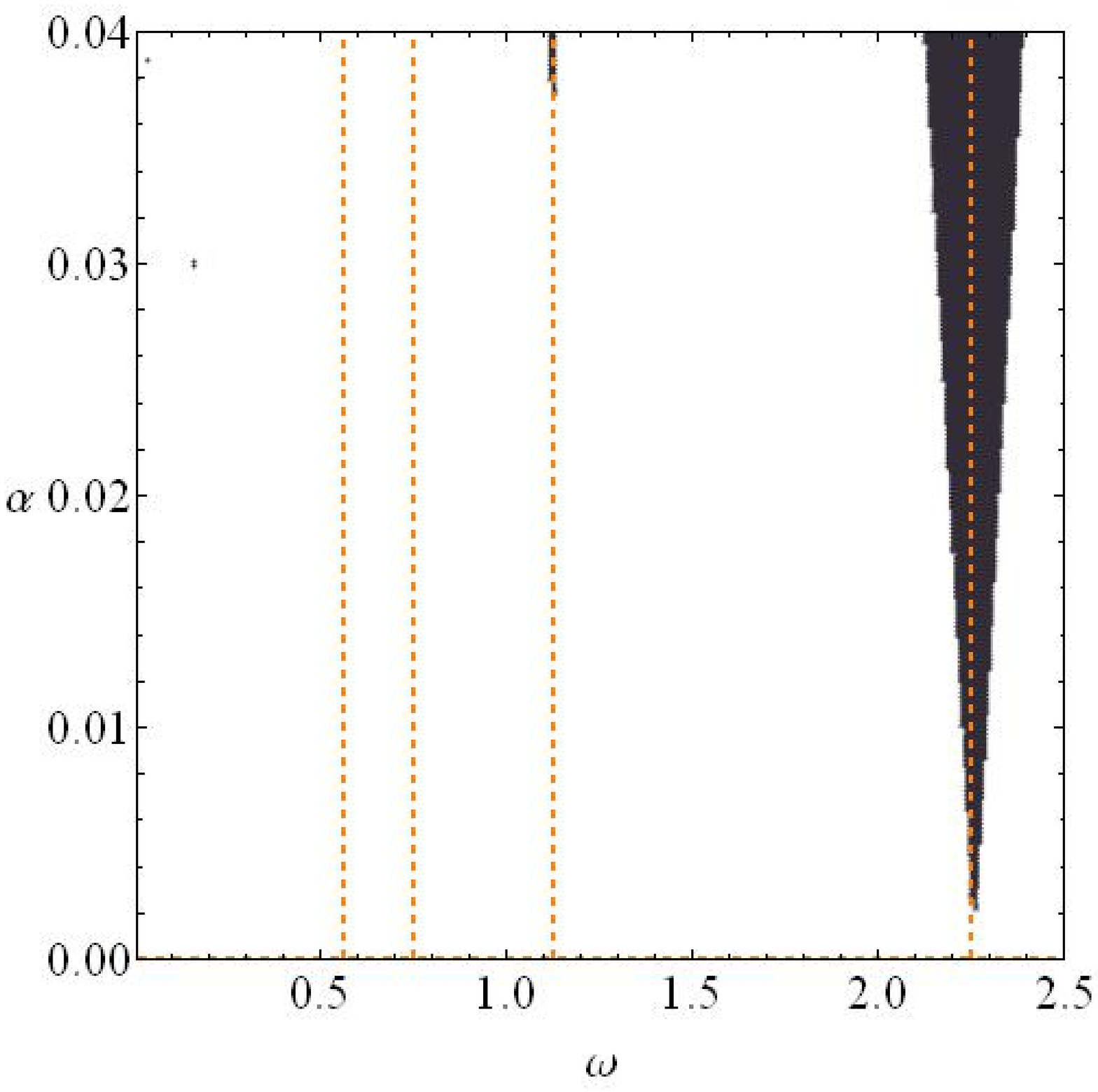}\quad
\includegraphics[width=4cm,height=4cm,clip]{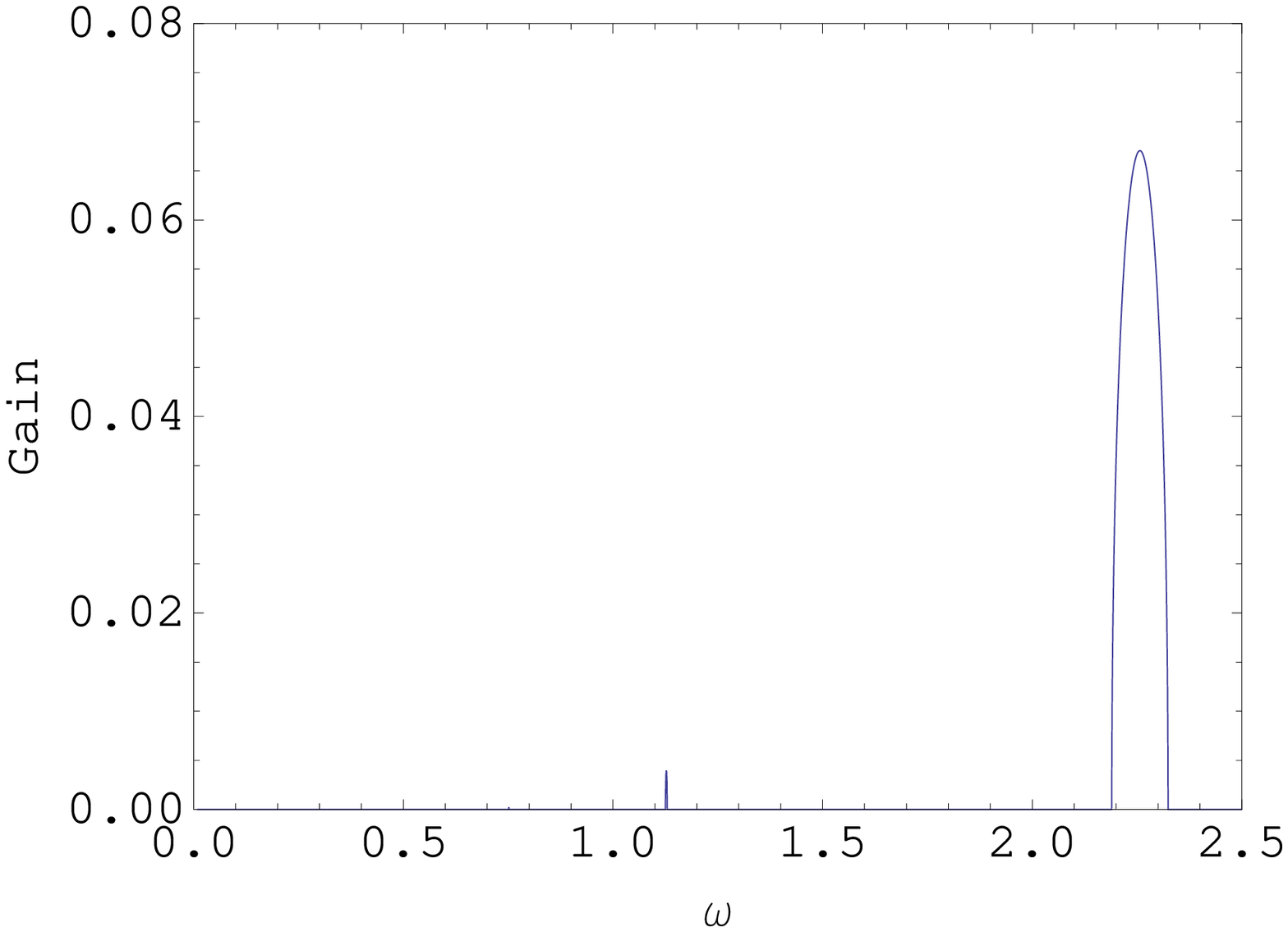}}
\caption{(Color online) The domains of instability (a, c) and gain
factors (b, d) concerning generation of a density wave with
$k=k_{rot}$ for the rotonized (a,b) and roton-free (c,d)
condensates. Vertical dashed lines represent sub-harmonics of the
main modulation frequency $\omega/n$, with $\omega=0.82$ (a),
$\omega=2.25$ (c), $n=1,2,3,4$. Parameter values: $A=2$, $g_0=-1$,
$q=1$ (a,b), $q=0.8$ (c,d), $k_{rot}=2.35$. } \label{fig3}
\end{figure}

Figure \ref{fig3} illustrates the domains of instability and gain
factors, corresponding to generation of density waves with
$k=k_{rot}$ in a rotonized, and roton-free condensates. As can be
noted from comparing the upper and lower panels of this figure, for
the rotonized case the contribution of the main modulation frequency
and its sub-harmonics towards generation of density waves greatly
exceeds the similar contribution for the roton-free condensate. This
implies, that if the excitation spectrum of a dipolar BEC features a
local minimum at $k=k_{rot}$ (see Fig. \ref{fig1}a, red line), then
the density waves are excited much more effectively.

\begin{figure}[htb]
\centerline{{\large a)} \hspace{4cm} {\large b)}} \centerline{
\includegraphics[width=4cm,height=4cm,clip]{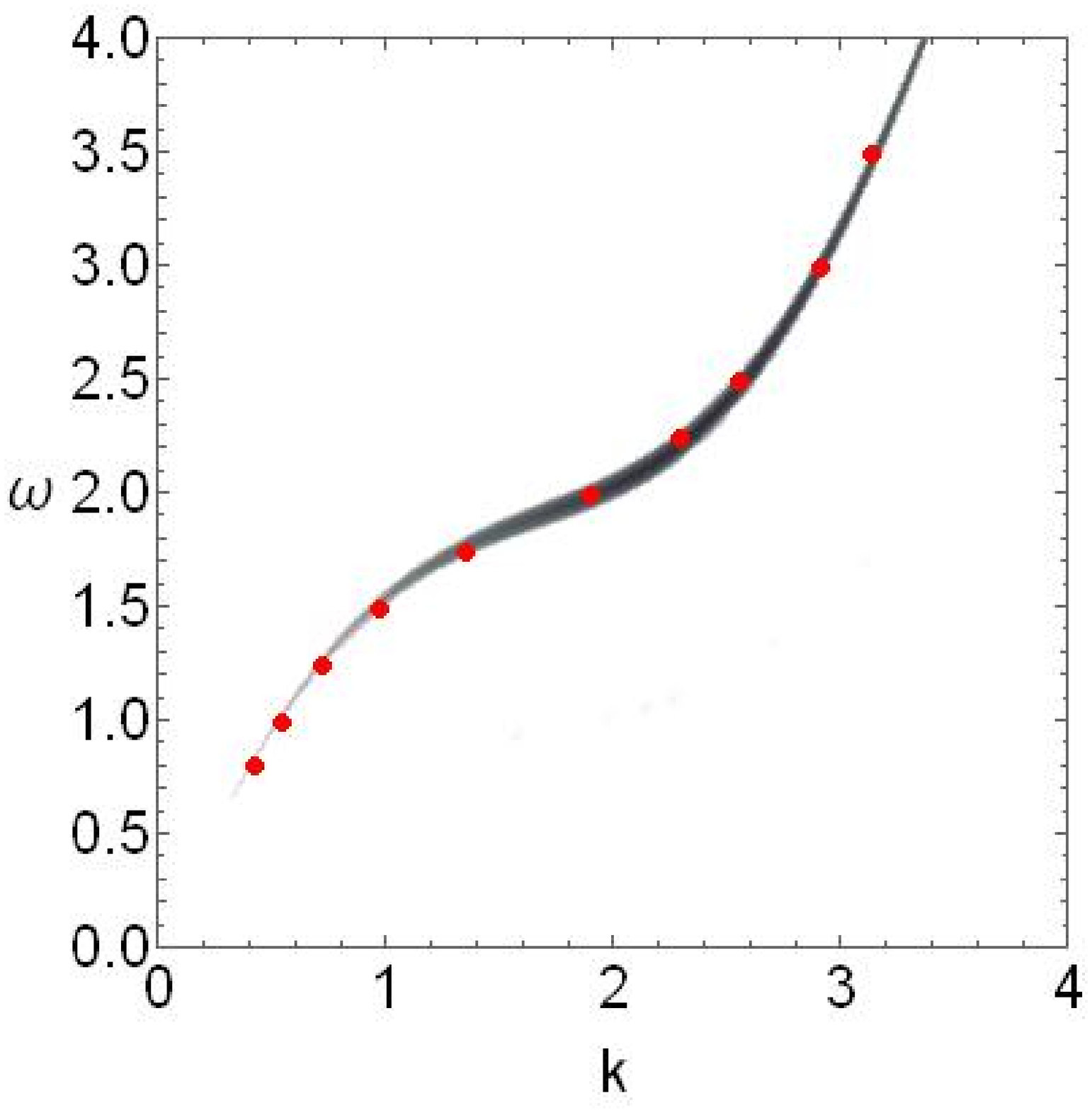}\qquad
\includegraphics[width=4cm,height=4cm,clip]{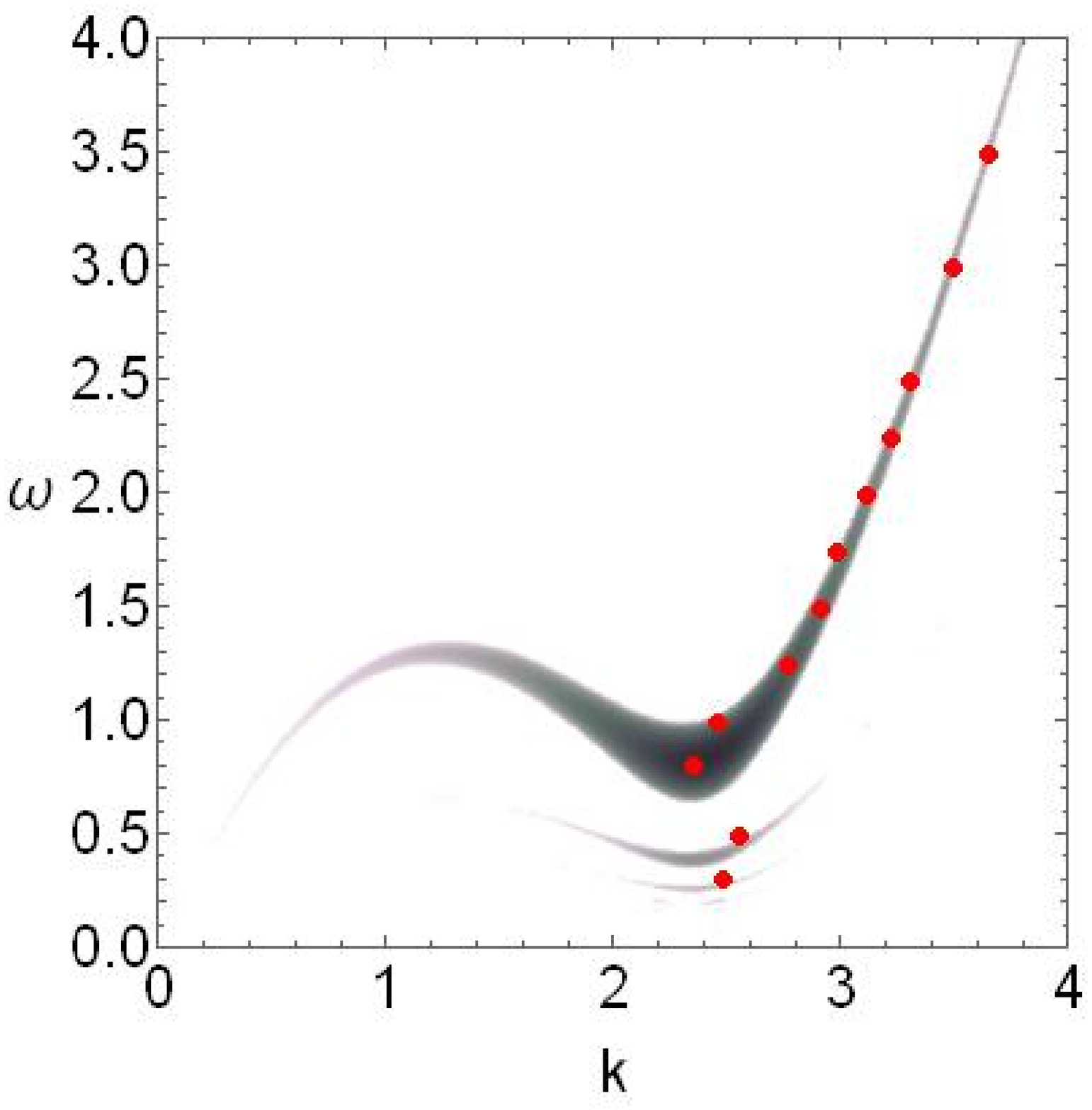}}
\caption{(Color online) The growth rate of instability $G(\omega,k)$
as a function of the modulation frequency and wave vector of
emerging waves, according to Floquet's theory (\ref{gain}),
represented through a density plot. The darker regions correspond to
higher gain factors~$G(\omega,k)$. Red symbols designate values
obtained from numerical simulations of Eq. (\ref{gpe}). The
parameters correspond to absence (a) and presence (b) of the roton
mode in the excitation spectrum. Parameter values: $A = 2$, $g_0 =
-1$, $p=0$, $\alpha =0.02$, $q=0.8$ (a), $q=1$ (b).} \label{fig4}
\end{figure}

In Fig. \ref{fig4} we show the gain factor as a function of two
input parameters: the frequency of modulations of the dipolar
interactions and wave vector of emerging density waves for the
roton-free and rotonized condensates.
In the same figures we show by symbols the result of numerical
simulations. To obtain these numerical data we solve the GPE
(\ref{gpe}) with time-periodic modulation of the strength of dipolar
interactions as per Eq. (\ref{gt}). When the amplitude of emerging
waves reaches few percent of the background amplitude, the
simulation is terminated and the spatial period of emerging density
waves is measured, and therefore its wave vector $k$ is determined.
In the roton-free case (Fig. \ref{fig4}a) the dependence $\omega(k)$
is monotonic, and the results of numerical simulations are in good
agreement with the prediction of the Floquet theory. For the same
reason, a good agreement is observed also in the rotonized case for
the frequency values above the local maximum, corresponding to wave
vectors $k>k_{rot}$. In the rotonized case (Fig.~\ref{fig4}b), for
one frequency value $\omega_0$ from the interval between the local
minimum and local maximum, there are three roots of the equation
$\omega(k)=\omega_0$. Despite the possibility of three resonant wave
vectors, the density wave with the greatest $k$ (out of three)
emerges. This can be understood by looking at Eq. (\ref{mathieu}),
where the strength of modulation is increasing function of the wave
vector $b \sim k^2 \bar{R}(k)$. For the values of modulation
frequency, below the roton minimum, the contribution of
sub-harmonics dominate (see Fig. \ref{fig3} a,b), resulting in
generation of density waves mainly with $k \simeq k_{rot}$. Thus, in
the condensate with a rotonized excitation spectrum, low frequency
modulations of the dipolar interactions gives rise to emergence of
density waves, corresponding to rotons.

\subsection{Generation of persistent density waves in dipolar condensates}

\begin{figure}[htb]
\centerline{{\large a)} \hspace{4cm} {\large b)}} \centerline{
\includegraphics[width=4cm,height=4cm,clip]{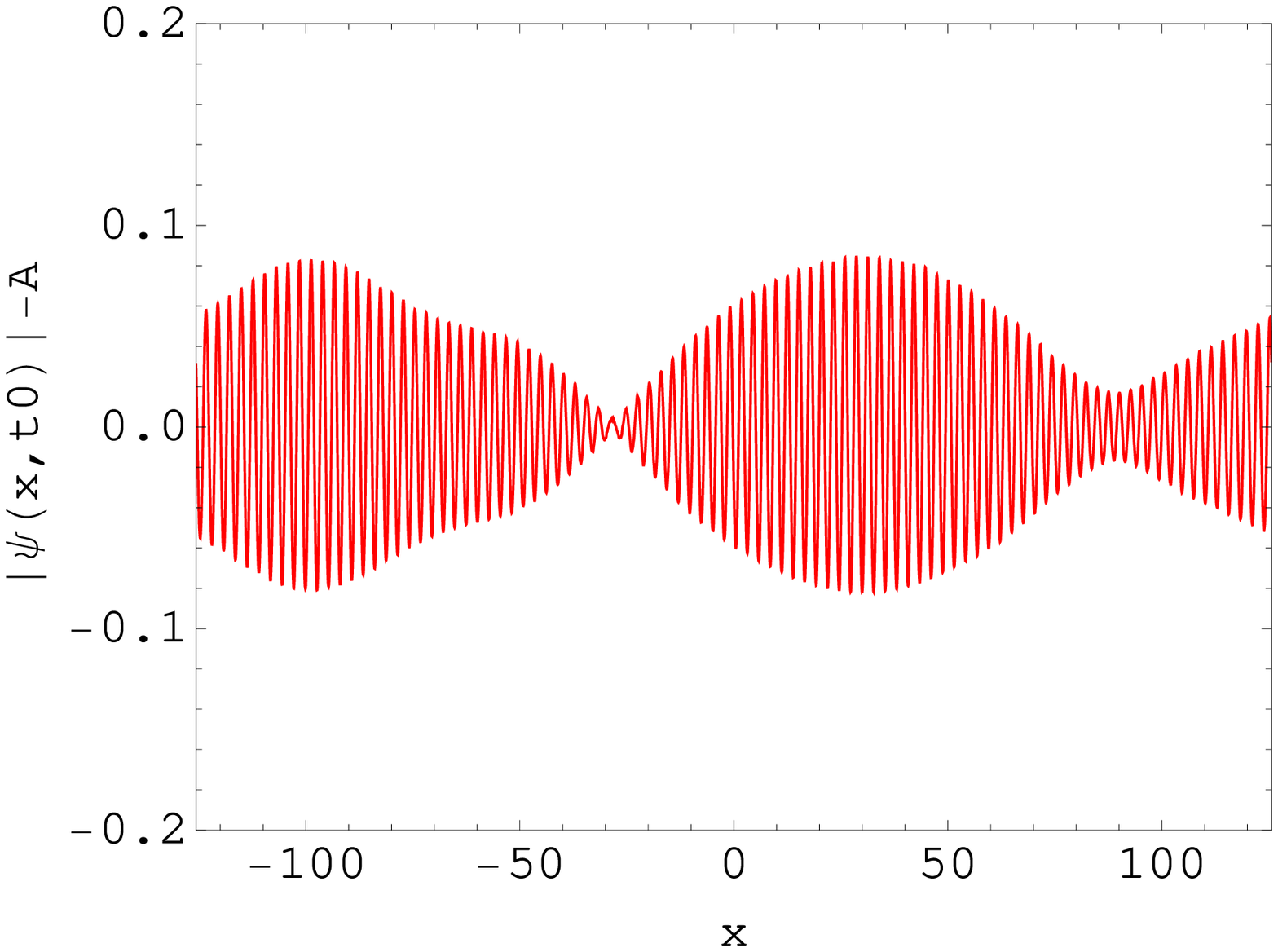}\qquad
\includegraphics[width=4cm,height=4cm,clip]{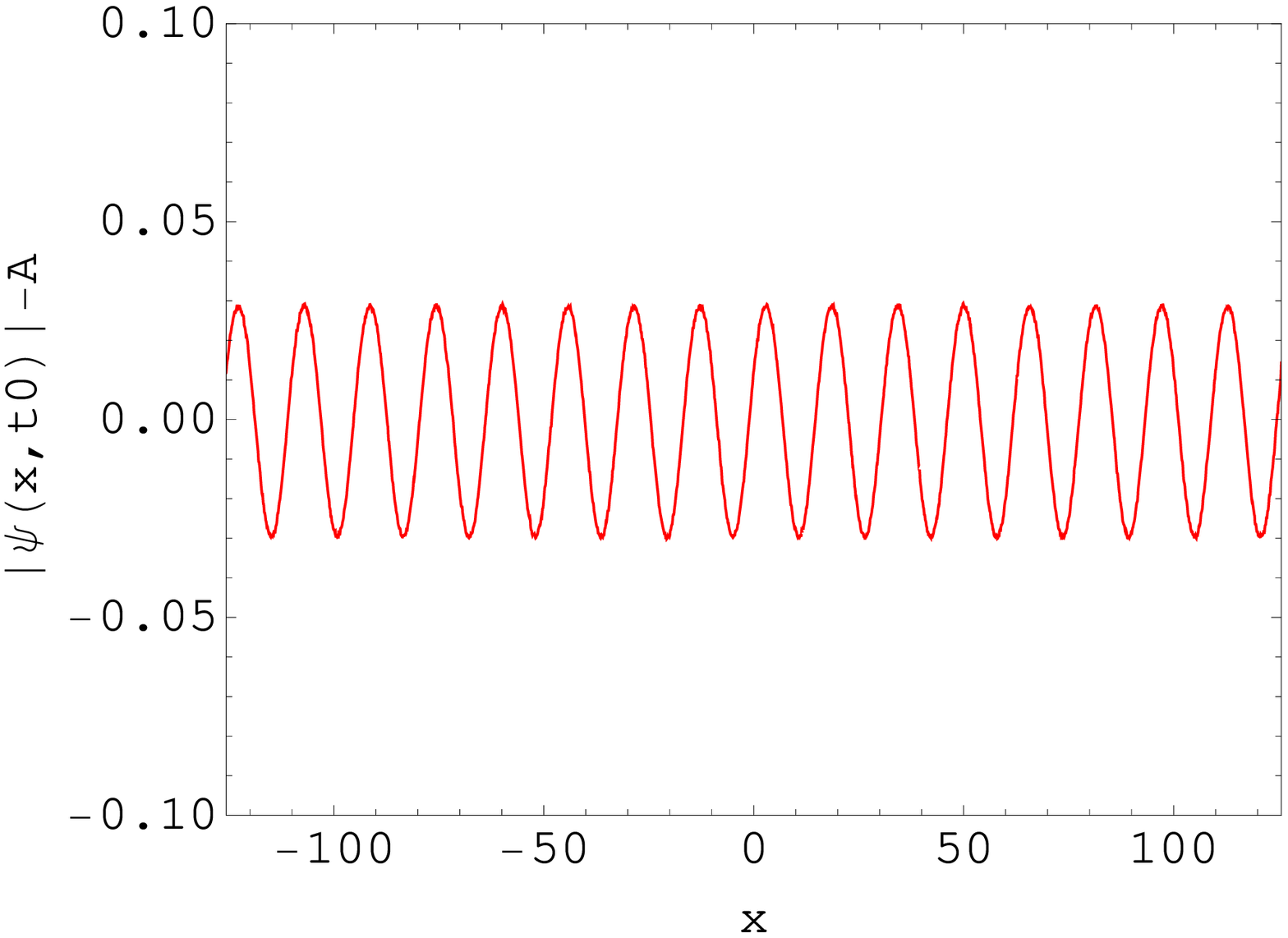}}
\caption{(Color online) Typical density waves in a dipolar BEC with
rotonized (a) and roton-free (b) excitation spectrum, emerging at
particular time instance $t_0=87$ (a) and $t_0=600$ (b), in
numerical simulations of the GPE (\ref{gpe}). The period of density
waves $\nu=2\,\pi/k$: $\nu=2.69$ (a), $\nu=15.23$ (b). Parameter
values: $A=2$, $q=1$ (a), $q=0.8$ (b), $g_0=-1$, $p=0$,
$\alpha=0.02$, $\omega=0.8$. } \label{fig5}
\end{figure}

From the analysis of previous sections it follows that time periodic
modulation of the strength of dipolar interactions gives rise to
spatially periodic density waves in the condensate. In those regions
of the parameter space, where the gain factor is a broad function of
the wave vector  $k \in [k_{min},k_{max}]$, as for example near the
roton minimum, the emerging density wave represents a superposition
of many wave components. Constructive and destructive interference
of these components produce a picture, where a high frequency wave
appears as modulated by a low frequency envelope (Fig.~\ref{fig5}a).
In the opposite case of a narrow, delta function-like gain factor
$G(k) \sim \delta(k)$, a wave with a single spatial period is
generated (Fig.~\ref{fig5}b).

\begin{figure}[htb]
\centerline{
\includegraphics[width=6cm,height=6cm,clip]{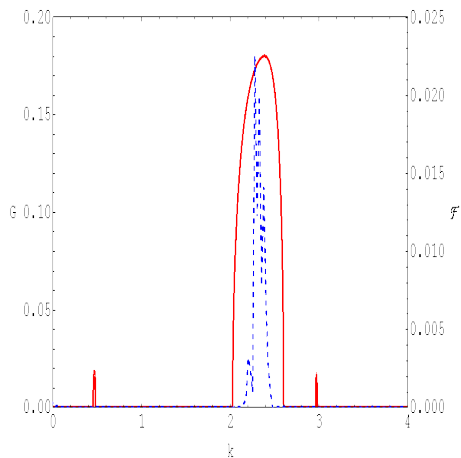}}
\caption{(Color online) Theoretical gain factor $G(k)$ (red solid
line, left axis) and Fourier transform ${\cal F}(k)$ (blue dashed
line, right axis) of the density wave, shown in Fig. \ref{fig5}a.}
\label{fig6}
\end{figure}

The Fig. \ref{fig6} illustrates the correlation between the
predictions of theoretical Floquet analysis for the gain factor, and
the results of direct numerical GPE simulations with additional
Fourier spectral analysis. A good qualitative agreement between the
theoretical approach and GPE simulations, observed in this figure,
confirms the adequacy of the developed model.

\begin{figure}[htb]
\centerline{\includegraphics[width=6cm,height=6cm,clip]{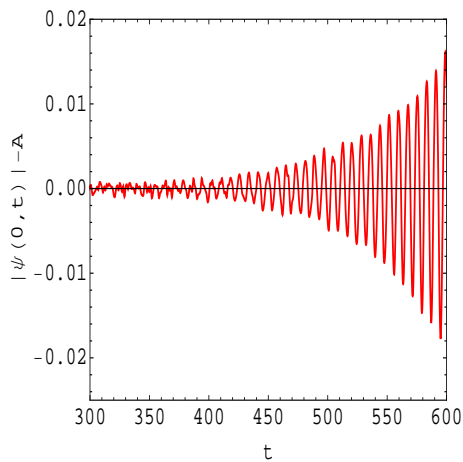}}
\caption{(Color online) The amplitude of the density wave at origin
resonantly increases with time, according to numerical solution of
Eq.~(\ref{gpe}). The temporal frequency, counted from this figure is
$\omega_{0} \simeq 0.79$, which is the half of the driving frequency
$2\,\omega=1.6$ in Eq. (\ref{gt}), suggesting the presence of the
parametric resonance phenomenon. Parameter values are similar to
those in Fig. \ref{fig5}b. } \label{fig7}
\end{figure}

The density waves, considered in this work are the consequence of
resonance phenomena triggered by time-periodic modulation of the
system parameters. Hence, the amplitude of waves steadily grows with
time, as depicted in Fig. \ref{fig7} obtained by numerical solution
of the governing Eq. (\ref{gpe}). The frequency of oscillations,
determined from this figure, is equal to half of the driving
frequency in Eq. (\ref{gt}), which is the evidence of the parametric
resonance phenomenon~\cite{landau}.
The main characteristic feature of Faraday waves is that, they
emerge at the modulation frequency, corresponding to parametric
resonance \cite{cross1993,nguyen2019}).

\begin{figure}[htb]
\centerline{{\large \qquad a)} \hspace{6cm} {\large b)} }
\centerline{
\includegraphics[width=6cm,height=8cm,clip]{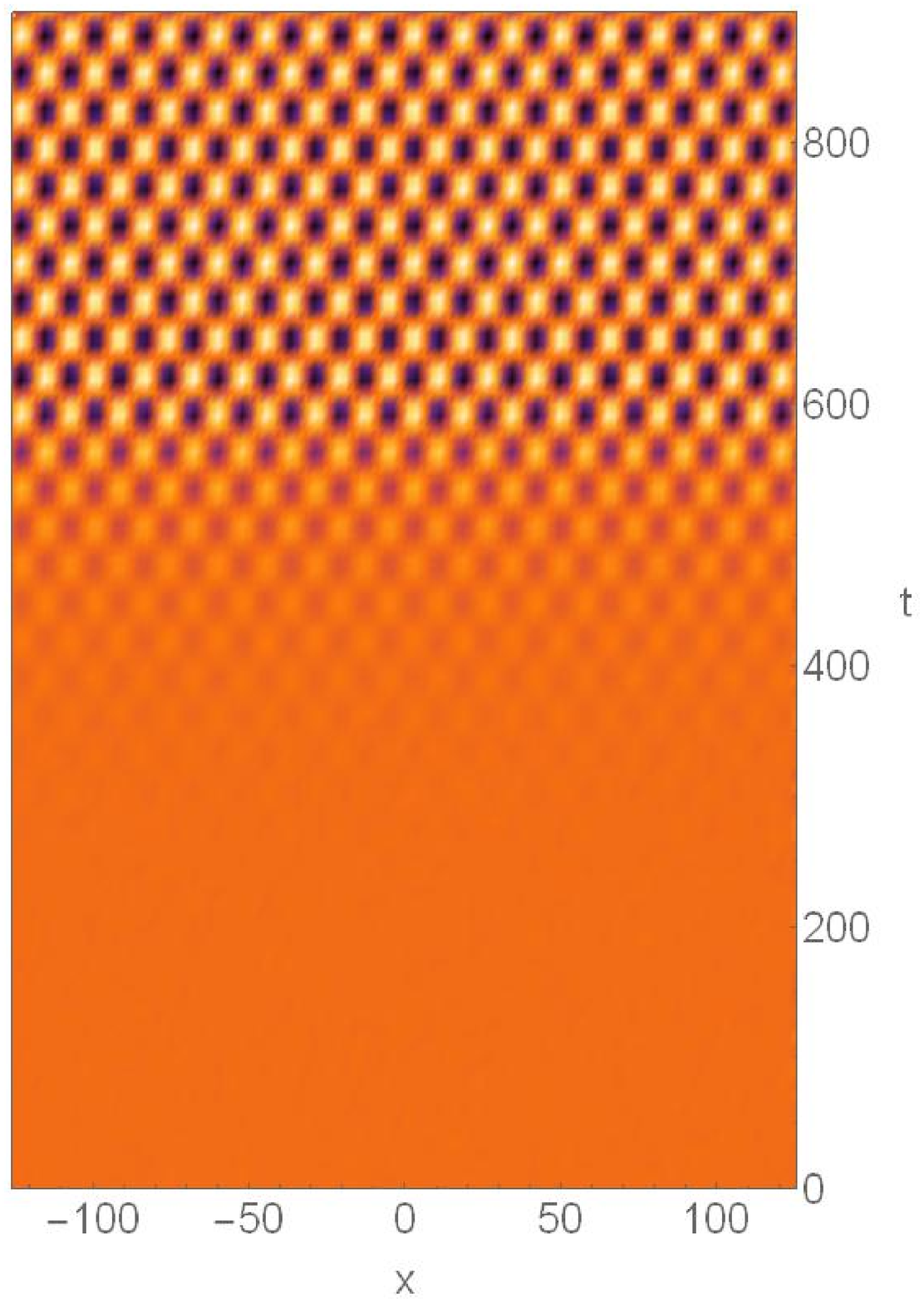}\qquad
\includegraphics[width=4cm,height=8cm,clip]{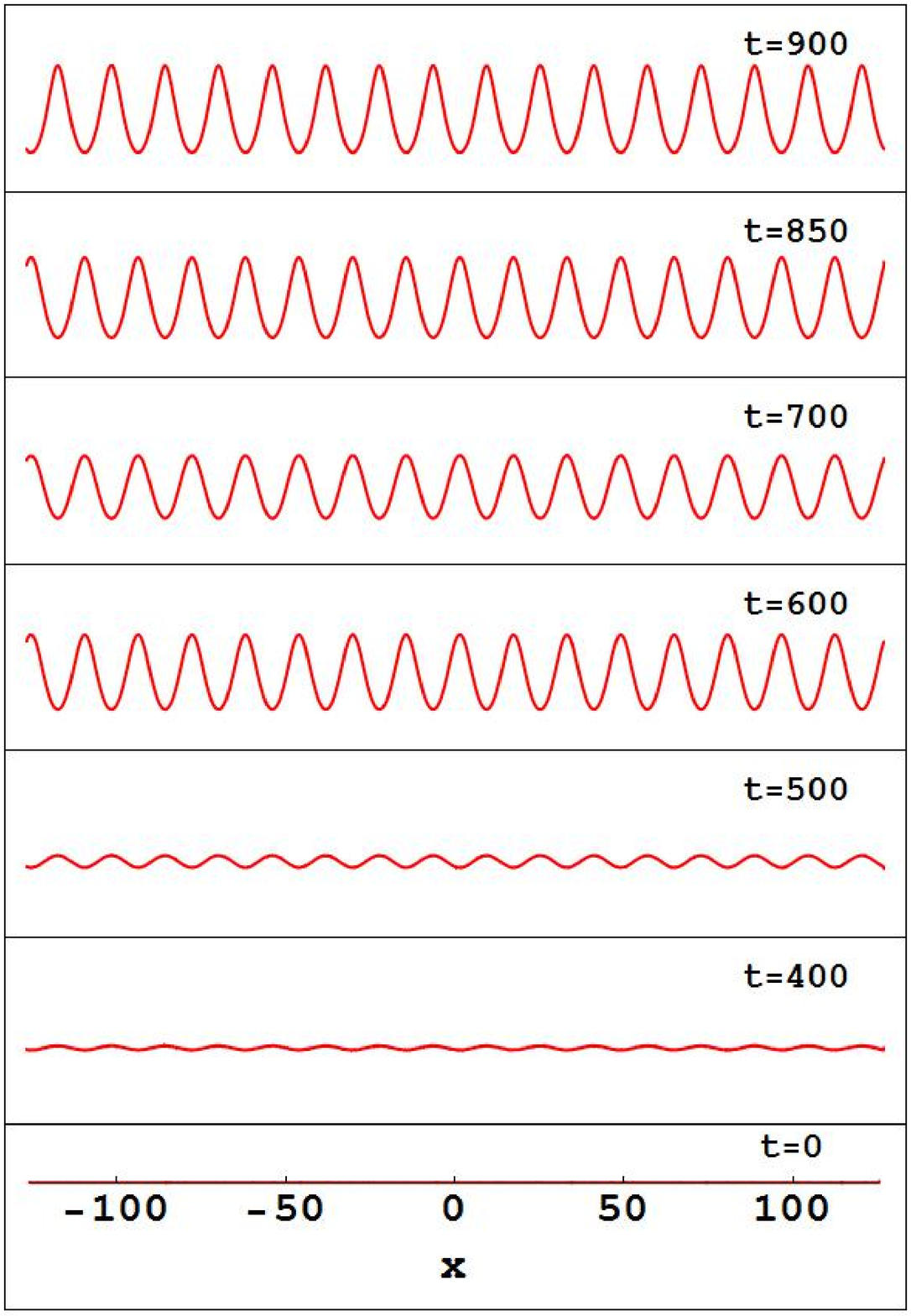}}
\caption{(Color online) Creation of persistent density waves by
time-periodic variation of the strength of dipolar interactions
according to Eq.~(\ref{gt}) with $\alpha(t)=\alpha_0 \cdot
\Theta(t_0-t)$, where $\Theta(t)$ is the Heaviside step function.
The results are obtained by numerical solution of the nonlocal GPE
(\ref{gpe}) and represented through a density plot (a) and 1D
waveforms at particular time sections (b). Parameter values: $A=2$,
$q=0.8$, $g_0=-1$, $p=0$, $\alpha_0=0.02$, $\omega=0.8$, $t_0=600$.
} \label{fig8}
\end{figure}

An interesting possibility would be to generate a steady wave in the
condensate, which can persist after the creation. To this end we
performed numerical experiments in which the periodic modulation in
Eq. (\ref{gt}) was kept until some time instance $t_0$, after that
was set to zero, i.e. $\alpha(t) = \alpha_0 \Theta(t_0-t)$ with
$\Theta(t)$ being the Heaviside theta function. The result for the
condensate with a roton-free excitation spectrum is shown in Fig.
\ref{fig8}. As can be seen in this figure, the density waves persist
afterwards the creation, provided that the modulation of the
coefficient of dipolar interactions is stopped at some time instance
$t \sim 600$. In fact this is expected outcome, since the original
model GPE~(\ref{gpe}) is conservative.

\begin{figure}[htb]
\centerline{{\large a)} \hspace{6cm} {\large b) \qquad \qquad}}
\centerline{
\includegraphics[width=4cm,height=4cm,clip]{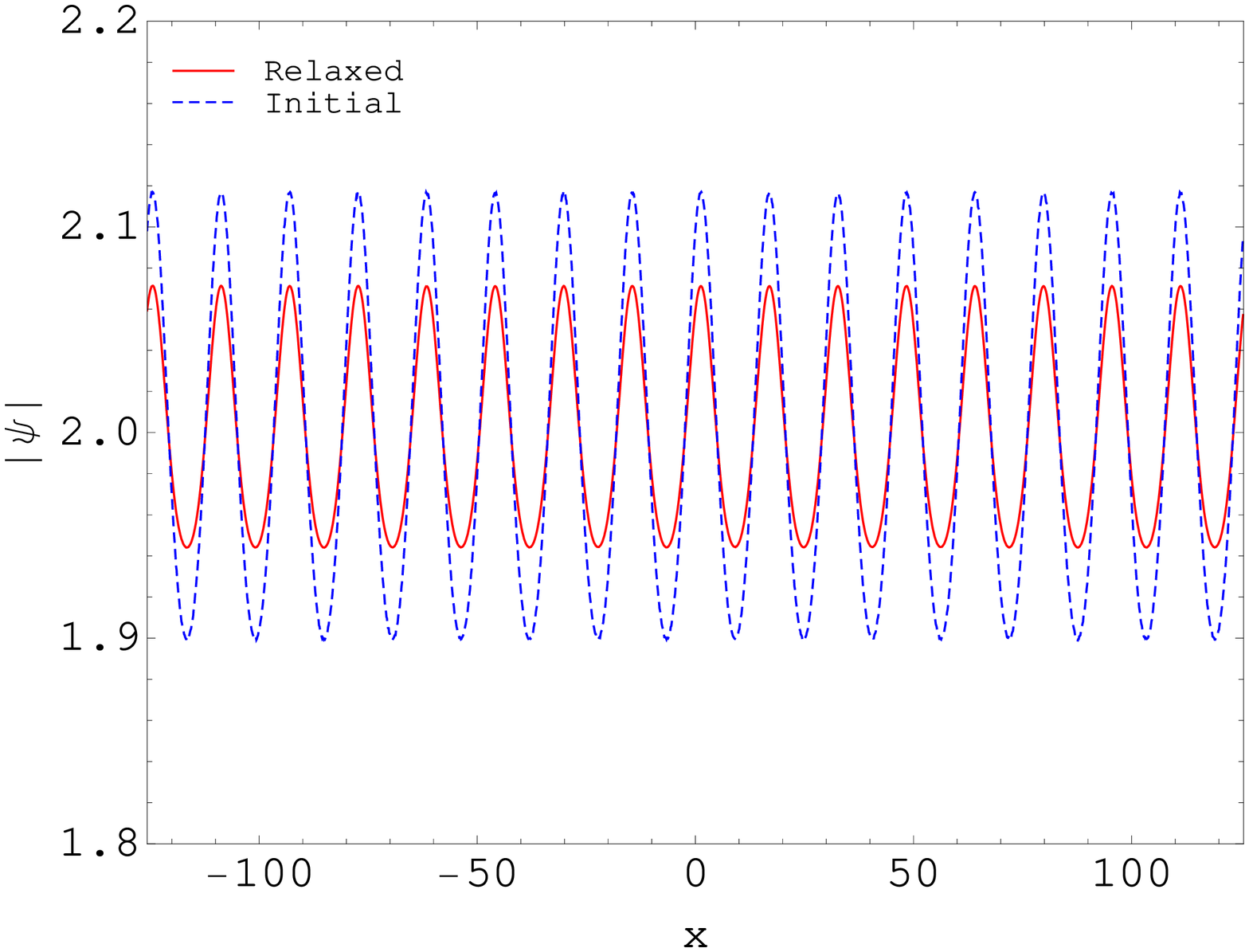}\qquad
\includegraphics[width=8cm,height=4cm,clip]{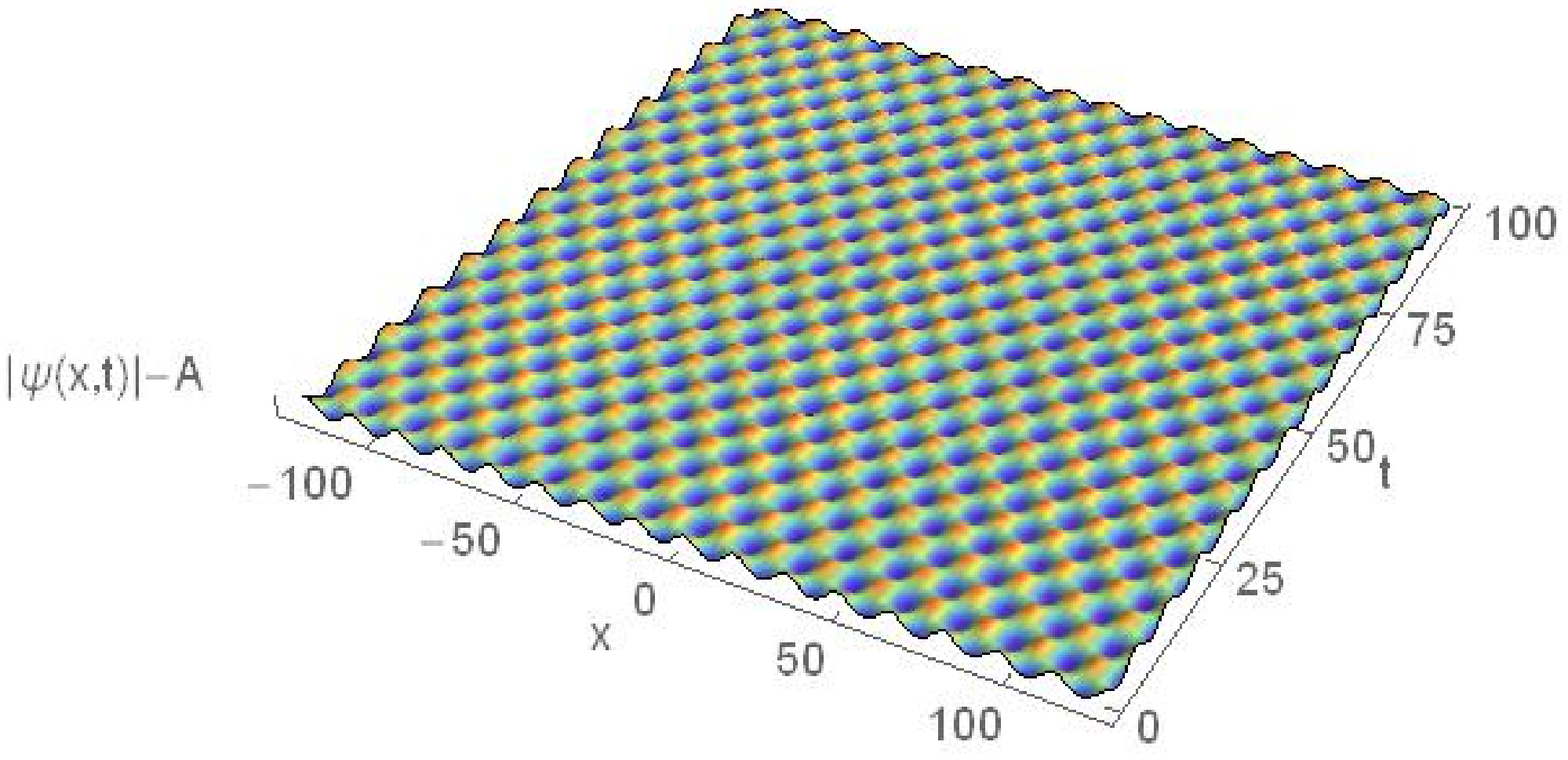}}
\caption{(Color online) (a) The density wave of Fig. \ref{fig8}b at
$t=600$ (blue dashed line), subjected to relaxation procedure,
transforms to a stationary periodic wave (red solid line). (b) When
the relaxed wave function is inserted as initial condition to GPE
(\ref{gpe}) and propagated in time, a standing wave pattern is
observed.} \label{fig9}
\end{figure}

In addition, the result of Fig. \ref{fig8} suggests, that GPE
(\ref{gpe}) may have a stationary spatially periodic solution. To
verify this conjecture, we pick the density wave at some particular
time instance, namely at $t=600$, and subject it to relaxation
procedure. Iterative relaxation method of Nijhof \cite{nijhof2000}
yields the result, shown in Fig. \ref{fig9}a. Next, by inserting the
relaxed wave-function as initial condition to GPE (\ref{gpe}) and
propagating in time,  we obtain the stationary wave pattern,
depicted in Fig. \ref{fig9}b. Thus we get a numerical evidence for
the existence of a stationary spatially periodic solution of the GPE
(\ref{gpe}).

In real experimental settings some dissipation effects may be
present, leading to damping of density waves. However, these issues
are outside of the scope of present work.

\subsection{The effect of three-body atomic interactions}

The density waves, which are under consideration in this work are
the consequence of nonlinear properties of the host medium. In the
original experiment \cite{engels2007}, the strength of radial
confinement of BEC in a cigar-shaped trap was periodically varied in
time, which resulted in generation of Faraday waves in the
longitudinal direction. This would not be possible in a system,
described by a linear Schr\"odinger equation, because without the
nonlinearity the radial perturbation would not be transferred to the
axial direction in the form of Faraday waves. Therefore, one can
expect enhancement of density waves in condensates with higher order
nonlinearities. A well known example is the quintic nonlinearity
originating from three-body atomic interactions in the condensate,
whose strength is designated by $p$ in Eq. (\ref{gpe}). The explicit
form of this coefficient in terms of material parameters was derived
in Ref. \cite{kohler2002}.

\begin{figure}[htb]
\centerline{{\large a)} \hspace{4cm} {\large b)}} \centerline{
\includegraphics[width=4cm,height=4cm,clip]{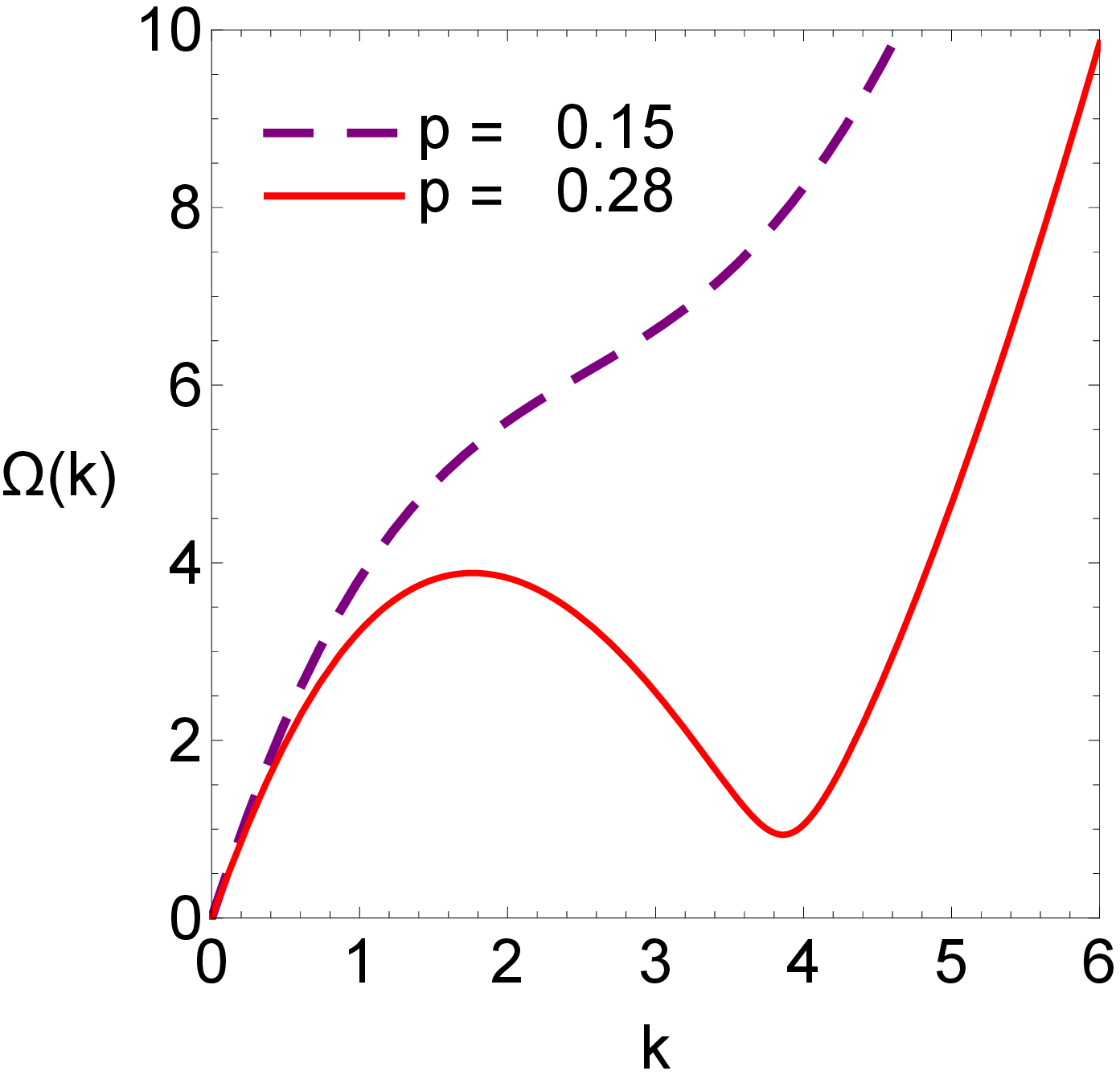}\qquad
\includegraphics[width=4cm,height=4cm,clip]{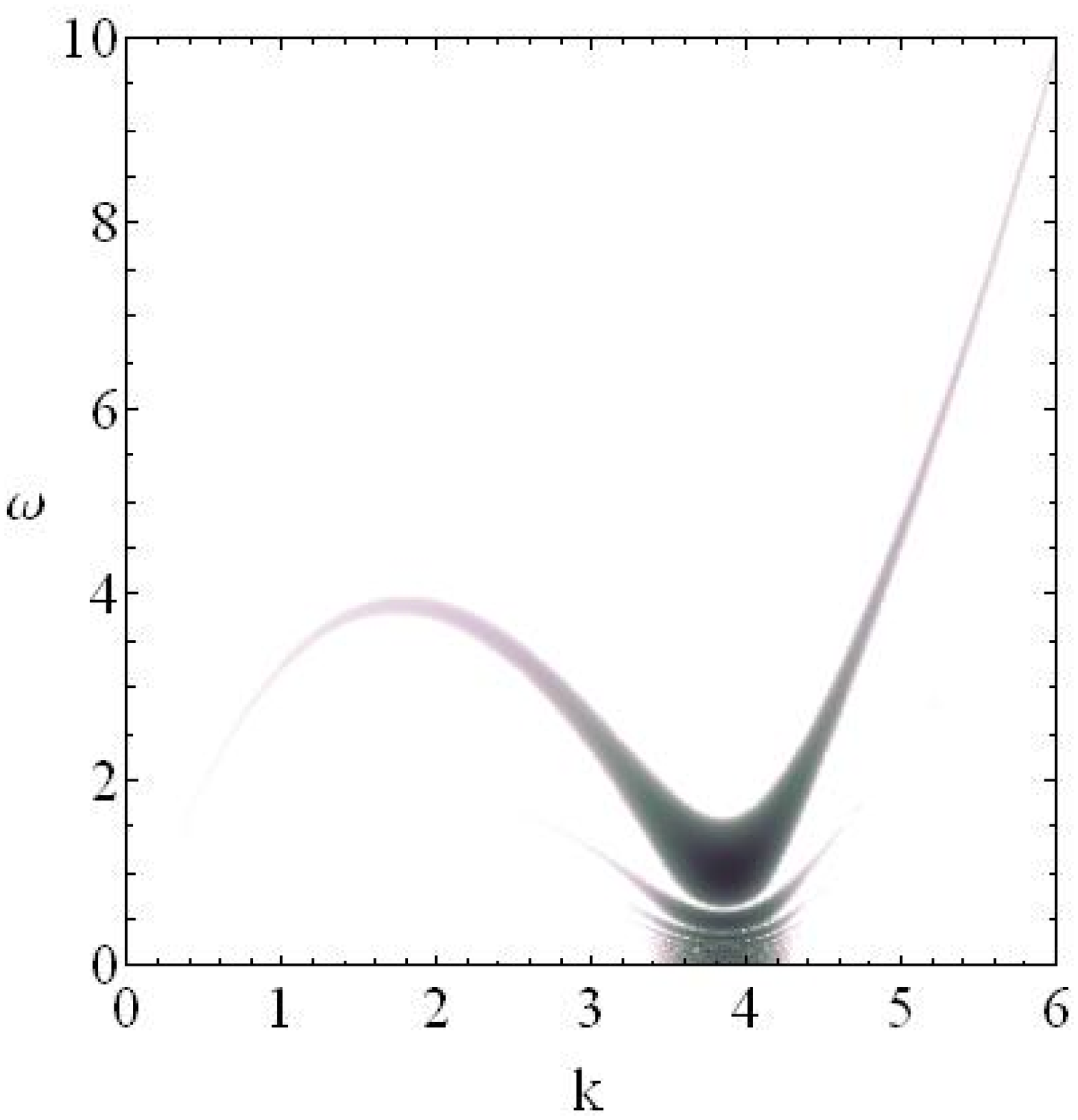}}
\caption{(Color online) The excitation spectrum (a) and gain factor
(b) as a function of the modulation frequency ($\omega$) and wave
vector ($k$), obtained from the Floquet theory (darker regions
correspond to higher gain factors). The roton minimum appears at
$k=3.87$. Parameter values: $A=2$, $q=0$, $g_0=-3.6$, $p=0.28$,
$\alpha=0.02$. } \label{fig10}
\end{figure}

\begin{figure}[htb]
\centerline{{\large a)} \hspace{4cm} {\large b)}} \centerline{
\includegraphics[width=4cm,height=4cm,clip]{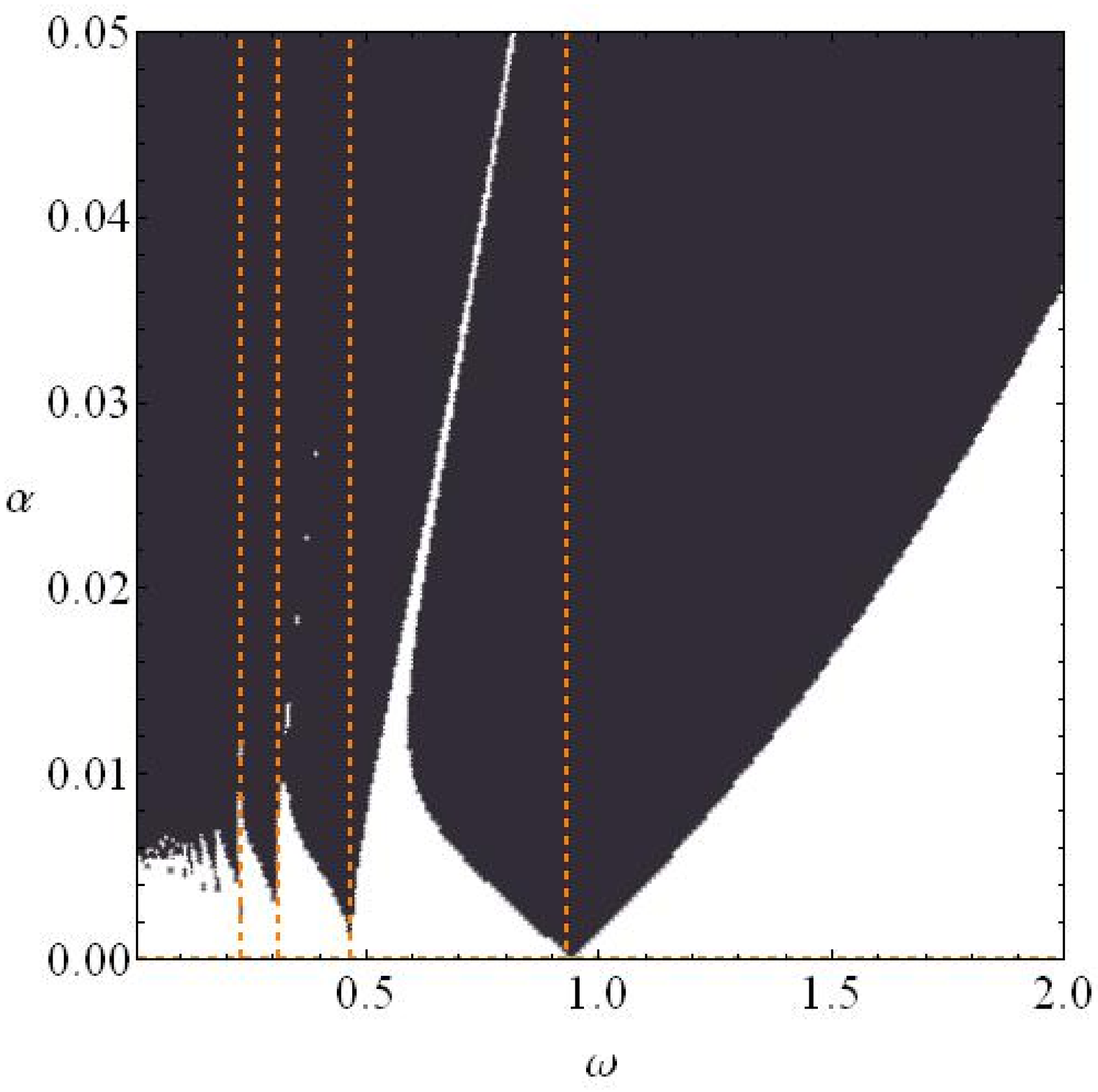}\qquad
\includegraphics[width=4cm,height=4cm,clip]{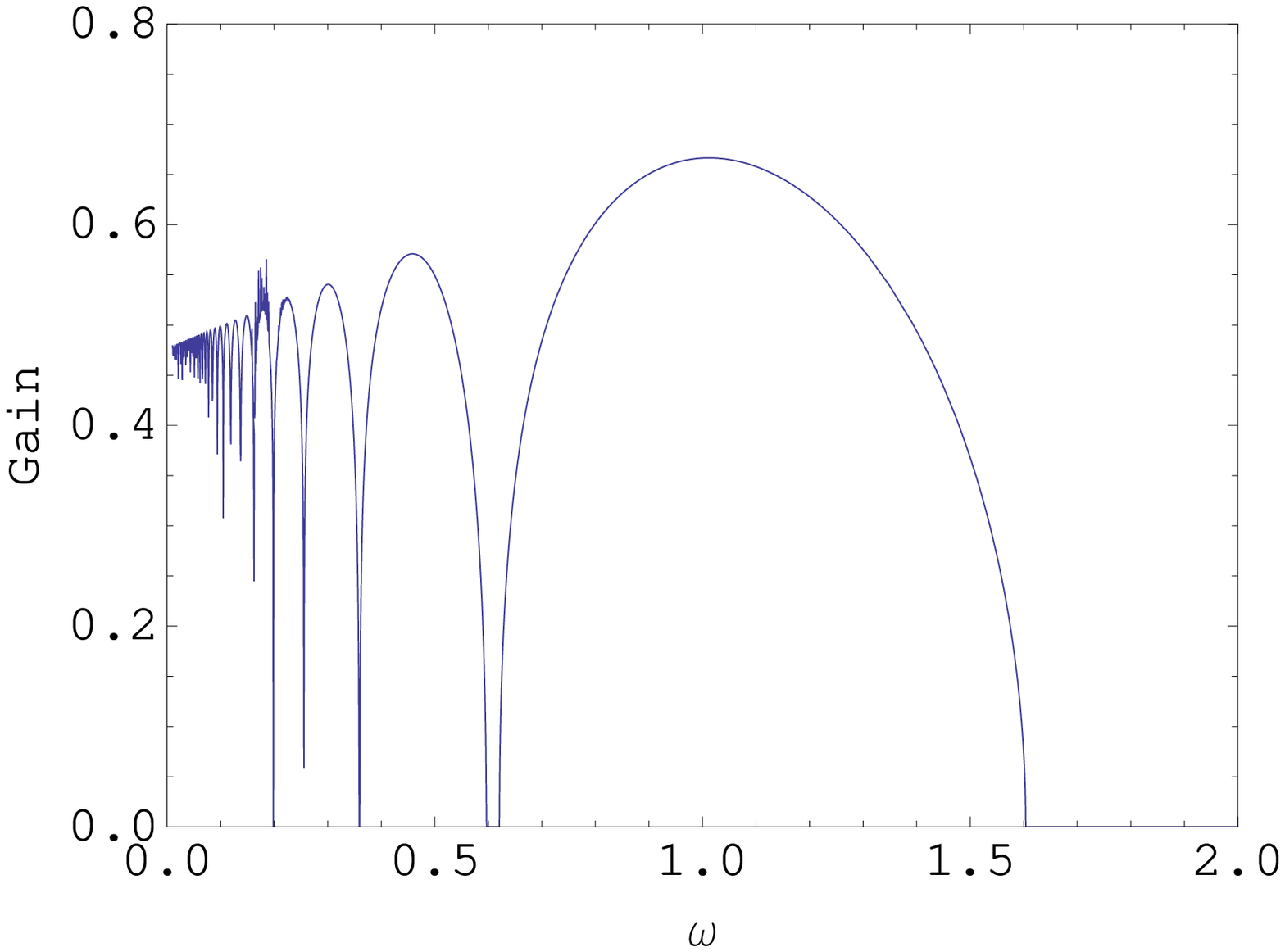}}
\caption{(Color online) (a) The domains of instability (dark
regions) for the GPE (\ref{gpe}) in the parameter space ($\alpha$,
$\omega$). (b) The gain factor $G(\omega)$ corresponding to
excitation of density waves near the roton minimum. Nearly equal
contribution of the main frequency $\omega=0.93$ and its
sub-harmonics $\omega/n$, $n=2,3,4...$ to generation of density
waves can be observed. Parameter values: $A=2$, $q=0$, $g_0=-3.6$,
$p=0.28$, $k=3.87$.} \label{fig11}
\end{figure}

In discussions about BECs with cubic and cubic-quintic
nonlinearities, it is usually assumed that two- and three-body
atomic interactions are relatively weak. However, in some
situations, when the inter-atomic coupling is moderate or strong,
models with only quintic nonlinearity (without the cubic term in Eq.
(\ref{gpe}), so $q=0$) are most accurate \cite{kolomeisky2000,
baizakov2009}. In Fig. \ref{fig10} we show the excitation spectrum
Eq. (\ref{Omega2}) and growth rate of instability according to
Floquet theory Eq. (\ref{gain}) for the condensate with only quintic
and dipolar atomic interactions.
By comparing this figure with its counterpart possessing only cubic
and dipolar interactions, shown in Fig. \ref{fig1}a, one notices a
strong shift of the roton minimum to greater values of the wave
vector. Besides, the contribution of sub-harmonics of the main
frequency towards generation of density waves with $k \simeq
k_{rot}$ appears to be much more pronounced in this case
(Fig.~\ref{fig10}b, Fig.~\ref{fig11}).

Figure \ref{fig11} illustrates the domains of instability as a
function of the strength and frequency of modulation in
Eq.~(\ref{gt}), and gain factor for the condensate with only quintic
and dipolar interactions.
From the Floquet analysis and numerical simulations of this section
one can conclude, that the higher order nonlinearity greatly
enhances the production of density waves in dipolar BECs.

\subsection{Experimentally relevant parameters}

To estimate experimentally relevant parameters we consider a BEC of
$^{164}$Dy atoms, whose magnetic dipole moment, background $s$-wave
scattering length and atomic mass are $d= 10 \,\mu_{B}$, $a_s=100 \,
a_0$, $m_{Dy}=2.72 \times 10^{-25} \, kg$, respectively, with
$\mu_{B}$, $a_0$ being the Bohr magneton and Bohr radius. The
condensate of $N=7 \times 10^4$ atoms is held in a tight quasi-1D
trap with radial confinement frequency $\omega_{\bot}=2 \pi \times
150$~Hz. The corresponding radial harmonic oscillator length, which
is the adopted length scale in this work, is $a_{\bot}=0.64 \,
\mu$m. The ratio between the strengths of dipolar and contact
interactions, computed with above defined parameters, is found to be
$\epsilon_d=a_{dd}/a_s~\simeq~1.31$, therefore in the ground state
we have the dipolar interaction dominated regime. The coefficients
of two-body contact interactions and long-range dipolar interactions
in physical units are $g_1=-2.74 \times 10^{-51} \, kg \cdot
m^5/s^2$ and $C_{d} = 1.08 \times 10^{-50} \, kg \cdot m^5/s^2$,
yielding the dimensionless parameter $g_0=C_{d}/g_1 \simeq - 3.9$.
The strength of three-body atomic interactions $K_3=K_r+i\, K_{i}$
in $^{164}$Dy condensate was reported to be $K_r=5.87 \times
10^{-39} \, \hbar \cdot m^6/s$ (real, conservative part) and
$K_{i}=7.8 \times 10^{-42} \, \hbar \cdot m^6/s$ (imaginary part,
characterizing the three-body recombination rate)~\cite{bisset2015}.
Since the imaginary part of this parameter is smaller than its real
part by three orders of magnitude, we have omitted it in the GPE
(\ref{gpe}), leaving only the conservative part $g_2=K_r$. The
coefficient of quintic nonlinearity, computed using this value and
above defined parameters is found to be $p \simeq 0.01$. Numerical
simulations are performed using the integration domain of length
$L=80 \, \pi$, which corresponds in physical units to ${\cal L} = L
a_{\bot} \simeq \, 162 \, \mu$m. The amplitude of the background
wave in dimensionless units is defined as $A=\sqrt{2\,a_s\,N/{\cal
L}} \simeq 2.14$, where $N$ stands for the number of atoms in the
condensate.

The above estimates of dimensionless parameters $A$, $q$, $g_0$ and
$p$ derived from experimentally feasible quantities are in the range
of values, used in our numerical simulations. Some deviations can be
adjusted by changing the tunable parameters of the system, such as
the $s$-wave scattering length, frequency of radial confinement and
the strength of dipolar interactions.

\section{Conclusions}

We have shown, that time-periodic variation of the strength of
atomic interactions gives rise to spatially periodic density waves
in dipolar quantum gases. The features of emerging density waves are
found to be quite different when the roton mode in the excitation
spectrum is present, compared to a roton-free case. In the former
case the emerging density waves show up as a superposition of waves
with different spatial periods, which interfere with each-other and
produce a beating-like patterns. In the latter case the density
modulations are close to a monochromatic wave. In numerical
experiments we observed persistent density waves, which can be
created by keeping the time-periodic modulation of the dipolar
interaction until the waves emerge, and afterwards setting the
modulation part to zero. By subjecting the emergent density wave to
a relaxation procedure, and subsequently evolving thus obtained wave
function with GPE, we observed a standing wave pattern. The
existence of a stationary spatially periodic solution of the
nonlocal GPE is conjectured. Theoretical analysis, based on the
instability of small perturbations of the background density allowed
to evaluate the gain factor for arbitrary set of parameters in the
governing nonlocal GPE. The quintic nonlinearity is found to greatly
enhance the production of density waves in dipolar BECs. The domains
of instability in the parameter space of the problem have been
identified, and theoretical predictions are confirmed by direct
numerical GPE simulations.

\acknowledgments

We thank E. N. Tsoy for useful discussions and helpful advises. This
work has been supported by grant $\Phi$A-$\Phi$2-004 of the Ministry
of Innovative Development of Uzbekistan.

\end{document}